\documentclass[a4paper,11pt]{article}
\pdfoutput=1
\usepackage{jheppub}
\usepackage[T1]{fontenc}
\def\lsim{\mathrel{\rlap{\lower4pt\hbox{\hskip1pt$\sim$}}
    \raise1pt\hbox{$<$}}}    
\def\gsim{\mathrel{\rlap{\lower4pt\hbox{\hskip1pt$\sim$}}
    \raise1pt\hbox{$>$}}}                

\newcommand{\beq}{\begin{eqnarray}}
\newcommand{\eeq}{\end{eqnarray}}

\newcommand{\tr}{{\rm Tr}}

\preprint{MAN/HEP/2012/021}
\title  {125 GeV Higgs Boson and the Type-II Seesaw Model}
\author[a]{P. S. Bhupal Dev,}
\author[b]{Dilip Kumar Ghosh,}
\author[c]{Nobuchika Okada}
\author[b]{and Ipsita Saha}
\affiliation[a]{Consortium for Fundamental Physics, School of Physics and Astronomy,\\ University of Manchester, Manchester, M13 9PL, United Kingdom}
\affiliation[b]{Department of Theoretical Physics,
Indian Association for the Cultivation of Science,\\
2A \& 2B Raja S.C. Mullick Road, Kolkata 700032, India}
\affiliation[c]{Department of Physics and Astronomy, University of Alabama, Tuscaloosa, AL 35487, USA}
\emailAdd{Bhupal.Dev@hep.manchester.ac.uk}
\emailAdd{tpdkg@iacs.res.in}
\emailAdd{okadan@ua.edu}
\emailAdd{tpis@iacs.res.in}
\abstract{We study the vacuum stability and unitarity conditions for a 125 GeV Standard Model (SM)-like Higgs boson mass in the type-II seesaw model. We find that, as long as the seesaw scale is introduced below the SM vacuum instability bound, there exists a large parameter space predicting a 125 GeV Higgs mass, irrespective of the exact value of the seesaw scale, satisfying both stability and unitarity conditions up to the Planck scale. 
We also study the model predictions for the Higgs partial decay widths in the diphoton and $Z$+photon channels with respect to their SM expectations and find that the decay rates for these two processes are correlated. We further show that for any given enhancement in the 
Higgs-to-diphoton rate over its SM expectation, there exists an {\it upper} bound on the type-II seesaw scale, and hence, on the masses of the associated doubly- and singly-charged Higgs bosons in the allowed parameter space. For instance, if more than 10\% enhancement persists in the Higgs-to-diphoton channel, the upper limit on the type-II seesaw scale is about 450 GeV which is completely within the reach of the 14 TeV LHC. We believe this to be an encouraging result for the experimental searches of the singly- and doubly-charged Higgs bosons which, in combination with improved sensitivity in the Higgs-to-diphoton and Higgs-to-$Z$+photon channels, could probe the {\it entire} allowed parameter space of the minimal type-II seesaw model, and establish/eliminate it as a single viable extension of the SM.}
\begin{document}
\maketitle
\section{Introduction}\label{1}
A new neutral boson with mass around 125 GeV has recently been observed with 
more than $5\sigma$ significance by both ATLAS~\cite{atlas} and CMS~\cite{cms} 
experiments at the LHC, which is also consistent with the earlier 
observations of an excess in the 115 - 140 GeV range made by the Tevatron~\cite{tev} experiment. While this new particle mostly resembles the highly sought after Standard Model (SM) Higgs boson ($h$), there still exist some deviations from the SM expectations of its signal strength in some decay modes~\cite{atlas-dec12,cms-dec12}, most notably in one of the highest mass resolution channels, namely, $h\to \gamma\gamma$~\cite{atlas-gg,cms-gg}. Although more data and further detailed analysis are required to confirm whether these deviations are just statistical fluctuations or indeed hints of some New Physics beyond the SM,  it might be worthwhile examining some of the beyond SM scenarios which could possibly lead to these deviations. 

Even if the newly discovered particle turns out to be {\it the} SM Higgs boson, its current favored mass range $M_h= 125\pm 1$ GeV~\cite{atlas-dec12,cms-dec12} will make the SM Higgs self-coupling negative in its Renormalization Group (RG) running at some energy scale below the natural Planck scale $M_P=1.2\times 10^{19}$ GeV, thus leading to an instability of the effective Higgs potential, if no New Physics is assumed at an intermediate scale 
(for a review, see e.g.,~\cite{review}). 
Recently, a next-to-next-to-leading order (NNLO) analysis of the SM Higgs potential derived the following lower bound on the Higgs boson mass from the 
condition of absolute vacuum stability up to the Planck scale~\cite{nnlo}: 
\begin{eqnarray}
M_h({\rm GeV}) > 129.4 + 1.4\left[\frac{M_t({\rm GeV})-173.1}{0.7}\right]-0.5\left[\frac{\alpha_3(M_Z)-0.1184}{0.0007}\right]\pm 1.0 ({\rm th}). 
\end{eqnarray}
According to this result, the vacuum stability of the SM up to the Planck scale is excluded at 98\% CL for 
$M_h<126$ GeV. Putting it another way, $M_h= 125\pm 1$ GeV leads to the vacuum instability scale $\Lambda_I=10^9$ - $10^{12}$ GeV (depending on the exact values of the top quark mass and the strong coupling constant) at which the Higgs self-coupling becomes negative, and hence, necessitates the existence of some new physics beyond the SM at or below $\Lambda_I$ in order to make the electroweak vacuum absolutely stable (or sufficiently long-lived compared to the age of the Universe) all the way up to the Planck scale~\footnote{The issue of electroweak vacuum stability for a SM Higgs mass in the vicinity of 125 GeV is not yet fully 
settled since this mass range is close to the transition between absolute stability and metastability~\cite{nnlo, Ellis:2009tp, lindner, strumia, shapo, alekhin, masina}, and the uncertainties (both theoretical and experimental), 
mainly on the top mass, are too large at the moment. In addition, the 
effective quantum field-theoretic treatment and the RG evolution of the SM parameters may not be valid all the way up to the Planck scale due to some 
non-perturbative quantum gravity effects. A proper treatment of all these yet unresolved issues is beyond the scope of this paper.}. 
This provides us one of the main motivations to consider a simple and testable 
extension of the SM that can alleviate the vacuum stability problem at high scales while being consistent with {\it all} the experimental results at low scales. 

On the other hand, the discovery of neutrino oscillations, and hence, non-zero neutrino masses from solar, atmospheric, reactor and accelerator neutrino experiments~\cite{nureview} has given us a definitive evidence of some beyond SM physics. A precise understanding of the smallness of neutrino masses, along with their observed large mixing is a potential gateway to New Physics~\cite{rabi-review}. It is certainly interesting to see if a simple extension of the SM which 
accounts for the non-zero neutrino masses and mixing can simultaneously solve the electroweak vacuum stability problem. Note that in general, the vacuum structure of the Higgs potential in any SM extension can be modified due to the addition of one or more of the following interaction terms to the SM Lagrangian: (i) Yukawa interactions associated with the generation of neutrino mass, (ii) additional Higgs interactions associated with the new degrees of freedom, and (iii) extended gauge interactions. 

The simplest theoretical way to obtain non-zero neutrino masses is by 
breaking the global $(B-L)$ symmetry of the SM. This can be parametrized within
 the SM through an effective dimension-5 operator due to 
Weinberg~\cite{weinberg}. There exist three tree-level realizations~\cite{ma} 
of this Weinberg operator using only renormalizable interactions, and are 
generically known as the {\it seesaw mechanism}. In the simplest case, the so-called 
type-I seesaw~\cite{type1a, type1b, type1c, type1d, type2b}, the SM particle content is supplemented by SM 
gauge-singlet right-handed (RH) Majorana neutrinos ($\nu_R$), thus leading to 
the effective Weinberg operator of the type 
$y_{ij}(L^{\sf T}_i\Phi)(L^{\sf T}_j\Phi)/M$, 
where $L=(\nu_\ell,\ell)_L^{\sf T}$ (with $\ell=e,\mu,\tau$) 
is the $SU(2)_L$ lepton doublet, $\Phi=(\phi^+,\phi^0)^{\sf T}$ is the 
SM Higgs doublet, and $M$ is the heavy neutrino mass scale. The presence of 
these massive RH neutrinos has an impact on the structure of the Higgs 
potential above the seesaw scale $M$, and hence on its vacuum stability, 
via loop corrections induced by the RH neutrino 
Yukawa couplings~\cite{casas, shafi1, strumia, chen, zhang, mohanty, masina}. 
The requirement that the electroweak vacuum has a lifetime longer than the age 
of the Universe implies an upper bound on the RH neutrino mass scale of
 $10^{13}-10^{14}$ GeV (depending on the physical masses of the light neutrinos) for $M_h=125$ GeV~\cite{casas, strumia, masina}. However, it was found that the vacuum stability lower bound on the SM Higgs mass {\it increases} for large Dirac neutrino Yukawa couplings, and for the Majorana neutrino mass in the LHC-accessible TeV range, a 125 GeV vacuum stability bound requires that the Dirac neutrino Yukawa coupling responsible for the light-heavy neutrino mixing should be small: $y < 0.1$~\cite{mohanty} thus limiting our ability to observe the heavy neutrino signals at the LHC. For current collider limits on the heavy Majorana neutrino mass and its mixing with the light neutrinos, see~\cite{atlas-majorana,cms-majorana}.          

Another realization of the seesaw mechanism, known as the type-II seesaw~\cite{type2a, type2b, type2c, type2d, type2e}, is by adding an $SU(2)_L$ triplet scalar field $\vec{\Delta}=(\Delta^{++},\Delta^+,\Delta^0)$ carrying hypercharge $Y=2$ to the SM, thus leading to the effective Weinberg operator $(L^{\sf T}\vec{\sigma}L)\cdot (\Phi^{\sf T}\vec{\sigma}\Phi)/M$, where $\sigma^i$'s are the usual $2\times 2$ Pauli matrices. One of the most interesting features of the type-II seesaw is that the seesaw messenger field $\Delta$, being a SM non-singlet, couples to the SM Higgs doublet via both cubic and quartic scalar couplings which has direct implications for the electroweak vacuum stability~\cite{shafi2, arhrib0, sahu, sharma, musolf, aoki, frank}. It turns out that a light SM Higgs boson in the (124 - 126) GeV range can easily be realized in type-II seesaw models. In fact, as shown in~\cite{shafi2}, the vacuum stability lower bound on the SM  Higgs boson mass is well below 125 GeV (even lower than the LEP2 Higgs mass bound of 114.4 GeV) for a TeV-scale type-II seesaw. Another important phenomenological consequence of the type-II seesaw mechanism is that the $SU(2)_L$ triplet scalar fields can  couple {\it directly} to the SM gauge bosons ($W^\pm,~Z,~\gamma$), and hence, are relatively easier to be detected at the LHC, if kinematically accessible~\cite{han,goran} as compared to the SM singlet heavy neutrinos 
in the type-I seesaw case which require a large mixing with the light neutrinos to have any observable effects at colliders. The current LHC bounds on the 
scalar triplet masses are given in~\cite{atlas-typeII,cms-typeII}.      

The third realization of the seesaw mechanism, known as the type-III seesaw~\cite{type3}, is by adding an $SU(2)_L$ fermion triplet $\vec{\Sigma}$, leading to the effective Weinberg operator $(L^{\sf T}\vec{\sigma}\Phi)^2/M$. The effect of these fermion triplets on the electroweak vacuum stability has been analyzed in~\cite{shafi1,shafi3}, and it was shown that similar to the type-II case, the vacuum stability bound becomes {\it lower} than the SM bound of 129 GeV for decreasing seesaw scale. These results have generated a lot of interest in type-II and III seesaw models~\footnote{The vacuum stability constraints for another variation of the seesaw, namely, linear and inverse seesaw models, have been recently studied in~\cite{sourov}.} in the light of the recent LHC discovery of the 125 GeV Higgs-like boson. For a review of various seesaw models and their testability at colliders and other experiments, see e.g.,~\cite{Nath:2010zj}.  

In this paper, we will focus on the minimal type-II seesaw model with a single $SU(2)_L$ scalar triplet added to the SM and analyze the parameter space allowed by the stability of the electroweak vacuum and the perturbativity of the scalar couplings for a 125 GeV SM-like Higgs mass. In this model, there are seven scalar mass eigenstates: two doubly-charged ($H^{\pm\pm}$), two singly-charged ($H^{\pm}$), two $CP$-even ($h,~H^0$) and a $CP$-odd ($A^0$) neutral scalar particles. However in a large part of the parameter space of this model, the mixing between the doublet and triplet scalar fields is usually small (unless the $CP$-even neutral scalar eigenstates are mass-degenerate), and hence, the lightest $CP$-even scalar field has essentially the same couplings to the SM fermions and massive vector bosons as the SM Higgs boson has~\cite{arhrib0, dey, akeroyd0, yagyu, yagyu1}. Therefore, we would expect its production rate as well as the branching ratios for its decay channels to be very similar to those of the SM Higgs boson, except for the loop-induced $h\to \gamma \gamma$~\cite{goran, arhrib1, carena} and also $h\to Z\gamma$~\cite{zg, frank, carena} channels which receive additional contributions from $H^{\pm\pm}$ and $H^\pm$ running in the loop. This has motivated many recent studies within the context of the Higgs triplet model~\cite{sharma, aoki, frank, goran, yagyu, am, ng, chiang} to explain the persistent `excess' in the $\gamma\gamma$ signal strength of the Higgs-like particle discovered at the LHC. 

Our approach in this paper is mainly motivated by the vacuum stability problem for a 125 GeV Higgs boson in the SM. As argued earlier, we require to have some new physics below the SM vacuum instability scale $\Lambda_I=10^9$ - $10^{12}$ GeV to solve this problem. We assume this new physics scale to be the type-II seesaw scale which is well-motivated for explaining the non-zero neutrino masses and mixing, and systematically analyze the modified RG running of the SM Higgs quartic coupling above this scale to obtain the allowed parameter space satisfying both vacuum stability and perturbativity conditions for all the couplings up to the Planck scale. We emphasize that there exists a large allowed parameter space for the type-II seesaw model which yields a 125 GeV SM-like Higgs boson pole mass, irrespective of the seesaw scale. We demonstrate this point by considering some typical values of the seesaw scale, both low (200 - 500 GeV) and high ($10^{9}$ - $10^{10}$ GeV). We have included the effects of the neutrino Yukawa couplings on the RG evolution of the scalar couplings, which cannot be neglected in certain cases, unlike what was previously assumed in the literature.

 As a consequence of the low-scale seesaw case, we also study the predictions for the $h\to \gamma \gamma$ and $h\to Z\gamma$ branching ratios in the allowed parameter space of the type-II seesaw model with respect to the purely 
SM expectations. We find a correlation between the $h\to \gamma \gamma$ and $h \to Z\gamma$ decay rates which, in combination with its collider signals, could be used to test the low-scale type-II seesaw model at 
the LHC, when the signal strength sensitivities in the $\gamma\gamma$ and $Z\gamma$ channels improve in future. We buttress the motivation for collider searches of the low-scale type-II seesaw by obtaining an 
{\it upper} bound on the seesaw scale, and hence, on the associated doubly- and singly-charged Higgs boson masses, for a given enhancement in the 
$h\to \gamma\gamma$ decay rate over its SM expectation, under the assumption that no other new physics effects contribute to this enhancement. For example, 
with an enhancement of 10\% or more, the corresponding upper limit on the type-II seesaw scale is about 450 GeV which is completely within the reach of the 14 TeV LHC. Thus, the LHC might be able to offer a definitive answer to the fate of the type-II seesaw model as a single viable extension of the SM up to the Planck scale.  

The plan of this paper is as follows: In Section~\ref{2}, we review the type-II seesaw model and its scalar sector. In Section~\ref{3}, we discuss the RG evolution of the Higgs quartic coupling and other scalar couplings in the model, including the effect of the neutrino Yukawa couplings.  In Section~\ref{4}, we present our results for the allowed parameter space at a given seesaw scale predicting a SM-like Higgs mass of 125 GeV while satisfying the vacuum stability as well as the unitarity conditions up to the Planck scale. In Section ~\ref{5}, we study the predictions of the decay rates for the processes $h\to \gamma\gamma$ and $h\to Z\gamma$ with respect to the SM expectations in the allowed parameter space of the type-II seesaw model, and show the correlation between the two decay rates. We also derive upper bounds on the seesaw scale for a given enhancement in the $h\to \gamma\gamma$ signal strength. Our conclusions are given in Section~\ref{6}. The matching conditions for the $\overline{\rm MS}$ and pole masses of the top-quark and the Higgs boson are collected in 
Appendix~\ref{A}. 
\section{Review of the Type-II Seesaw Model}\label{2}
In this section, we briefly review the minimal type-II seesaw model (for a detailed discussion, see e.g.,~\cite{acco}) where, in addition to the SM fields, a triplet scalar field $\Delta$ is introduced, which transforms as $({\bf 3},2)$ under the $SU(2)_L\times U(1)_Y$ gauge group:
\begin{eqnarray}
\Delta = \frac{\sigma^i}{\sqrt 2}\Delta_i = \left(\begin{array}{cc}
\delta^+/\sqrt 2 & \delta^{++}\\
\delta^0 & -\delta^+/\sqrt 2
\end{array}
\right),
\end{eqnarray}
with $\Delta_1=(\delta^{++}+\delta^0)/\sqrt 2,~\Delta_2=i(\delta^{++}-\delta^0)/\sqrt 2,~\Delta_3=\delta^+$. The Lagrangian for this model is given by 
\begin{eqnarray}
{\cal L} = {\cal L}_Y+{\cal L}_{\rm kinetic}-{\cal V}(\Phi,\Delta),
\label{lag}
\end{eqnarray}
where the relevant kinetic and Yukawa interaction terms are respectively
\begin{eqnarray}
{\cal L}_{\rm kinetic} &=& {\cal L}_{\rm kinetic}^{\rm SM}+{\rm Tr}\left[\left(D_\mu \Delta\right)^\dag \left(D^\mu\Delta\right)\right]\, ,\\
{\cal L}_Y &=& {\cal L}_Y^{\rm SM}-\frac{1}{\sqrt 2}\left(Y_\Delta\right)_{ij} L_i^{\sf T}Ci\sigma_2\Delta L_j+{\rm H.c.}\, .
\end{eqnarray}
Here $C$ is the Dirac charge conjugation matrix with respect to the Lorentz group, and 
\begin{eqnarray} 
D_\mu \Delta = \partial_\mu \Delta + i\frac{g}{2}[\sigma^a W_\mu^a,\Delta]+i\frac{g'}{2}B_\mu \Delta \qquad (a=1,2,3)
\end{eqnarray}
is the covariant derivative of the scalar triplet field, with the GUT-normalization for the electroweak couplings $g=g_2$ and $g'=\sqrt{3/5}g_1$.  

Following the notation of~\cite{schmidt}, we write the scalar potential in Eq.~(\ref{lag}) as~\footnote{The general form of the potential given in~\cite{arhrib1} can be recovered with a simple redefinition of the couplings: $\lambda\to \lambda/2,~(\lambda_1+\lambda_2)\to 2\lambda_2,~\lambda_2\to -2\lambda_3,~(\lambda_4+\lambda_5)\to \lambda_1,~\lambda_5\to -\lambda_4/2$, and using the identity 
$(\Phi^\dag \Phi){\rm Tr}(\Delta^\dag \Delta)=\Phi^\dag\{\Delta^\dag,\Delta\}\Phi$ which is valid for any traceless $2\times 2$ matrix $\Delta$.}  
\begin{eqnarray} 
{\cal V}(\Phi,\Delta) &=& -m_\Phi^2(\Phi^\dag \Phi)+\frac{\lambda}{2}(\Phi^\dag \Phi)^2+M^2_\Delta {\rm Tr}(\Delta ^\dag \Delta)+ \frac{\lambda_1}{2}\left[{\rm Tr}(\Delta ^\dag \Delta)\right]^2\nonumber\\
&& +\frac{\lambda_2}{2}\left(\left[{\rm Tr}(\Delta ^\dag \Delta)\right]^2-{\rm Tr}\left[(\Delta ^\dag \Delta)^2\right]\right)+\lambda_4(\Phi^\dag \Phi){\rm Tr}(\Delta ^\dag \Delta)+\lambda_5\Phi^\dag[\Delta^\dag,\Delta]\Phi\nonumber\\
&& +\left(\frac{\Lambda_6}{\sqrt 2}\Phi^{\sf T}i\sigma_2\Delta^\dag \Phi+{\rm H.c.}\right)\, .
\label{eq:Vpd}
\end{eqnarray}
The coupling constants $\lambda_i$ can be chosen to be real through a phase redefinition of the field $\Delta$. Also, we have chosen $m_\Phi^2>0$ in order to ensure the spontaneous symmetry breaking of the $SU(2)_L\times U(1)_Y$ gauge group to $U(1)_Q$ by a nonzero vacuum expectation value (vev) for the neutral component of the SM Higgs doublet, $\langle \phi^0\rangle = v/\sqrt 2$ with $v\simeq 246.2$ GeV while $M_\Delta^2$ can be of either sign.  Note that the last term in Eq.~(\ref{eq:Vpd}) is the {\it only} source of lepton number violation at the Lagrangian level before the spontaneous symmetry breaking.  
\subsection{Neutrino Masses and Mixing}
A non-zero vev for the Higgs doublet field $\Phi$ induces a tadpole term for the scalar triplet field $\Delta$ via the $\Lambda_6$ term in Eq.~(\ref{eq:Vpd}), thereby generating a nonzero vev for its neutral component, $\langle \delta^0\rangle = v_\Delta/\sqrt 2$, and breaking lepton number by two units. This results in the following Majorana mass matrix for the neutrinos: 
\begin{eqnarray}
(M_\nu)_{ij} = v_\Delta (Y_\Delta)_{ij}
\label{eq:neutrino}
\end{eqnarray}
The triplet vev contributes to the weakly interacting gauge boson masses at tree-level:
\begin{eqnarray}
M_W^2 = \frac{g^2}{2}(v^2+2v^2_\Delta),~~M_Z^2 = \frac{g^2}{2\cos^2\theta_W}(v^2+4v_\Delta^2),
\end{eqnarray}
thereby affecting the SM $\rho$-parameter:
\begin{eqnarray}
\rho \equiv \frac{M_W^2}{M_Z^2\cos^2\theta_W} = \frac{1+\frac{2v_\Delta^2}{v^2}}{1+\frac{4v_\Delta^2}{v^2}}.
\end{eqnarray}
The electroweak precision data constraints require the $\rho$-parameter to be very close to its SM value of unity: $\rho=1.0004^{+0.0003}_{-0.0004}$~\cite{pdg}. This requires 
\begin{eqnarray}
\frac{v_\Delta}{v} < 0.02 \, , \quad {\rm or} \quad 
v_\Delta \lsim 5~{\rm GeV}.
\label{rho} 
\end{eqnarray}
Hence, we will be working in the limit $v_\Delta\ll v$. 

The electroweak symmetry breaking (EWSB) conditions are obtained after minimizing the scalar potential given by Eq.~(\ref{eq:Vpd}):
\begin{eqnarray}
m_\Phi^2 &=& \frac{1}{2}\lambda v^2-\Lambda_6 v_\Delta + \frac{\lambda_4-\lambda_5}{2}v_\Delta^2,\label{eq:ewsb1}\\ 
M_\Delta^2 &=& \frac{1}{2}\frac{\Lambda_6 v^2}{v_\Delta}-\frac{1}{2}(\lambda_4-\lambda_5)v^2-\frac{1}{2}\lambda_1 v_\Delta^2.
\label{eq:ewsb2}
\end{eqnarray}
In the limit $v_\Delta \ll v$, we obtain from Eq.~(\ref{eq:ewsb2})
\begin{eqnarray}
v_\Delta = \frac{\Lambda_6 v^2}{2M_\Delta^2+v^2(\lambda_4-\lambda_5)}\; .
\label{eq:vdelta}
\end{eqnarray}
Note that for $M_\Delta \gg v$, Eq.~(\ref{eq:neutrino}) for the neutrino masses becomes 
\begin{eqnarray}
M_\nu\simeq \frac{\lambda_6 v^2}{2M_\Delta}Y_\Delta
\label{eq:seesaw}
\end{eqnarray} 
with the dimensionless parameter $\lambda_6\equiv \Lambda_6/M_\Delta$. Eq.~(\ref{eq:seesaw}) resembles a typical seesaw formula with $M_\nu\propto M_\Delta^{-1}$. 

In order to satisfy the low-energy neutrino oscillation data~\cite{pdg}, 
we fix the structure of the Yukawa coupling matrix $Y_\Delta$ as follows: From 
Eq.~(\ref{eq:neutrino}), we obtain
\begin{eqnarray}
Y_\Delta = \frac{M_\nu}{v_\Delta} = \frac{1}{v_\Delta}U^{\sf T}M_\nu^{\rm diag}U
\label{YD}
\end{eqnarray}
where $M_\nu^{\rm diag} = {\rm diag}(m_1,m_2,m_3)$ is the diagonal neutrino mass eigenvalue matrix. $U$ is the standard Pontecorvo-Maki-Nakagawa-Sakata (PMNS) mixing matrix, usually parametrized in terms of the three mixing angles $\theta_{12},\theta_{23},\theta_{13}$, and one Dirac ($\delta$) and two Majorana ($\alpha_1,\alpha_2$) $CP$ phases:  
\begin{eqnarray}  
U = \left(\begin{array}{ccc}
c_{12}c_{13} & s_{12}c_{13} & s_{13}e^{-i\delta}\\
-s_{12}c_{23}-c_{12}s_{23}s_{13}e^{i\delta} &
c_{12}c_{23}-s_{12}s_{23}s_{13}e^{i\delta} & s_{23}c_{13}\\ 
s_{12}s_{23}-c_{12}c_{23}s_{13}e^{i\delta} &
-c_{12}s_{23}-s_{12}c_{23}s_{13}e^{i\delta} & c_{23}c_{13} 
\end{array}\right)\times{\rm
  diag}(e^{i\alpha_1/2},e^{i\alpha_2/2},1)\; \nonumber,
\end{eqnarray}
where $c_{ij}\equiv \cos\theta_{ij},~s_{ij}\equiv \sin\theta_{ij}$. 
For illustration purposes, we assume the Majorana phases in the PMNS matrix to be zero, and also choose a normal hierarchy for the neutrino masses with $m_1=0$ so that $m_2=\sqrt{\Delta m^2_{\rm sol}}$ and $m_3=\sqrt{\Delta m^2_{\rm atm}}$. Using the central values of a recent global analysis of the 
3-neutrino oscillation data~\cite{global}:
\begin{eqnarray}
&&
\Delta m^2_{\rm sol}=  7.62\times 10^{-5}~{\rm eV}^2,~
\Delta m^2_{\rm atm} = 2.55\times 10^{-3}~{\rm eV}^2,~\nonumber\\
&&
\theta_{12}=34.4^\circ,~ \theta_{23}=40.8^\circ,\theta_{13}=9.0^\circ,
\delta=0.8\pi\; ,
\end{eqnarray} 
we obtain the following structure of the Yukawa coupling matrix:
\begin{eqnarray}
Y_\Delta = \frac{10^{-2}~{\rm eV}}{v_\Delta}\times \left(\begin{array}{ccc}
0.31-0.12i & -0.09+0.32i & -0.72+0.37i \\
-0.09+0.32i & 2.53+0.04i & 2.19+0.01i \\
-0.72+0.37i & 2.19+0.01i & 3.07-0.03i
\end{array}\right).
\label{Ydstruct}
\end{eqnarray} 

From Eq.~(\ref{Ydstruct}), we see that there are two extreme cases:\\ 
(i) Small Yukawa couplings which correspond to large $v_\Delta\lsim {\cal O}(\rm GeV)$, where the upper bound comes from the $\rho$-parameter constraint given by Eq.~(\ref{rho}). \\
(ii) Large Yukawa couplings $\left(Y_{\Delta}\right)_{ij} \sim {\cal O}(1)$ which corresponds to small $v_\Delta \gsim {\cal O}(10^{-2}~{\rm eV})$. Note that $v_\Delta$ cannot be arbitrarily small because of the naturalness consideration for the 
Yukawa couplings in Eq.~(\ref{Ydstruct}).  
\subsection{Scalar Masses and Mixing} 
Expanding the scalar fields $\phi^0$ and $\delta^0$ around their vevs, we obtain 10 real-valued field components:
\begin{eqnarray}
\Phi = \left(\begin{array}{cc}
\phi^+ \\ \frac{1}{\sqrt 2}(v+\phi+i\chi)\end{array}\right),
\qquad
\Delta = \left(\begin{array}{cc}
\frac{\delta^+}{\sqrt 2} & \delta^{++}\\
\frac{1}{\sqrt{2}}(v_\Delta+\delta+i\eta) & -\frac{\delta^+}{\sqrt 2}
\end{array}\right)
\end{eqnarray}
which, upon minimization of the scalar potential ${\cal V}(\Phi,\Delta)$ in Eq.~(\ref{eq:Vpd}) with respect to the vevs, yields a $10\times 10$ squared mass matrix for the scalars. There are seven physical massive eigenstates $H^{\pm\pm},H^\pm,h,H^0,A^0$ and three massless Goldstone bosons $G^\pm,G^0$ which are eaten up to give mass to the SM gauge bosons $W^\pm,Z$. The physical mass eigenvalues for the scalar sector are given by
\begin{eqnarray}
m^2_{H^{\pm\pm}} &=& M_\Delta^2+\frac{1}{2}(\lambda_4+\lambda_5)v^2+\frac{1}{2}(\lambda_1+\lambda_2)v_\Delta^2, \label{doub-mass}\\
m^2_{H^\pm} &=& \left(M_\Delta^2+\frac{1}{2}\lambda_4v^2+\frac{1}{2}\lambda_1v_\Delta^2\right)\left(1+\frac{2v^2_\Delta}{v^2}\right),\label{sing-mass}\\
m^2_{A^0} &=& \left(M_\Delta^2+\frac{1}{2}(\lambda_4-\lambda_5)v^2+\frac{1}{2}\lambda_1v_\Delta^2\right)\left(1+\frac{4v^2_\Delta}{v^2}\right),\\
m^2_h &=& \frac{1}{2}\left(A+C-\sqrt{(A-C)^2+4B^2}\right),\\
m^2_{H^0} &=& \frac{1}{2}\left(A+C+\sqrt{(A-C)^2+4B^2}\right),\\
{\rm with}~~
A = \lambda v^2,&&
B = -\frac{2v_\Delta}{v}\left(M^2_\Delta+\frac{1}{2}\lambda_1 v_\Delta^2\right),~~
C = M^2_\Delta+\frac{1}{2}(\lambda_4-\lambda_5)v^2+\frac{3}{2}\lambda_1 v_\Delta^2.\nonumber
\label{eq:scamass}
\end{eqnarray}
Note that among the two $CP$-even neutral Higgs bosons, $m_{H^0}>m_{h}$ is always 
satisfied. 

The mixing between the doublet and triplet scalar fields in the charged, $CP$-even and $CP$-odd scalar sectors are respectively given by 
\begin{eqnarray}
\left(\begin{array}{c}
G^\pm \\ H^\pm \end{array}\right) &=& \left(\begin{array}{cc}
\cos\beta' & \sin\beta'\\
-\sin\beta' & \cos\beta'
\end{array}\right)
\left(\begin{array}{c}
\phi^\pm \\ \delta^\pm \end{array}\right),\\
\left(\begin{array}{c} 
h \\ H^0 \end{array}\right) &=& \left(\begin{array}{cc}
\cos\alpha & \sin\alpha\\
-\sin\alpha & \cos\alpha
\end{array}\right)
\left(\begin{array}{c}
\phi \\ \delta \end{array}\right),\\
\left(\begin{array}{c}
G^0 \\ A^0 \end{array}\right) &=& \left(\begin{array}{cc}
\cos\beta & \sin\beta\\
-\sin\beta & \cos\beta
\end{array}\right)
\left(\begin{array}{c}
\chi \\ \eta \end{array}\right),
\end{eqnarray} 
where the mixing angles are given by
\begin{eqnarray}
\tan\beta' &=& \frac{\sqrt 2 v_\Delta}{v},\label{mix1}\\
\tan\beta &=& \frac{2v_\Delta}{v}\equiv \sqrt 2 \tan\beta',\label{mix2}\\
\tan{2\alpha} &=& \frac{2B}{A-C}=\frac{4v_\Delta}{v}\frac{M_\Delta^2+\frac{1}{2}\lambda_1 v_\Delta^2}{M_\Delta^2+\frac{1}{2}(\lambda_4-\lambda_5-2\lambda)v^2+\frac{3}{2}\lambda_1 v_\Delta^2}\label{mix3}.
\end{eqnarray} 
Thus in the limit $v_\Delta\ll v$, the mixing between the doublet and triplet scalars is usually small (unless the $CP$-even scalars $h$ and $H^0$ are close to being mass-degenerate). In this limit, the mass of the (dominantly doublet) lightest $CP$-even scalar is simply given by $m_h^2=\lambda v^2$ (as in the SM) independent of the mass scale $M_\Delta$, whereas the other (dominantly triplet) scalars have $M_\Delta$-dependent mass. The mass scale $M_\Delta$ will be simply referred to as the ``seesaw scale" for the rest of our paper.  
\subsection{Stability and Unitarity Conditions}\label{stab}
Here we summarize the constraints on the scalar potential of the type-II seesaw model given by Eq.~(\ref{eq:Vpd}) in order to ensure the stability of the electroweak vacuum and the preservation of the tree-level unitarity in various scattering processes. 

The necessary and sufficient conditions valid for all directions in field space to ensure that the scalar potential in Eq.~(\ref{eq:Vpd}) is bounded from below are given by~\cite{arhrib1}
\begin{eqnarray}
&&\lambda \geq 0,~~\lambda_1\geq 0,~~2\lambda_1+\lambda_2 \geq 0,\nonumber \\
&&\lambda_4+\lambda_5+\sqrt{\lambda\lambda_1} \geq 0,~~
\lambda_4+\lambda_5+\sqrt{\lambda\left(\lambda_1+\frac{\lambda_2}{2}\right)} \geq 0,\nonumber \\
&&\lambda_4-\lambda_5+ \sqrt{\lambda\lambda_1} \geq 0,~~
\lambda_4-\lambda_5+\sqrt{\lambda\left(\lambda_1+\frac{\lambda_2}{2}\right)} \geq 0.
\label{eq:lstab}
\end{eqnarray} 
In addition, the tree-level unitarity of the $S$-matrix for elastic scattering imposes the following constraints~\cite{arhrib1}:
\begin{eqnarray}
&&\lambda \leq \frac{8}{3}\pi,~~
\lambda_1-\lambda_2\leq 8\pi,~~
4\lambda_1+\lambda_2\leq 8\pi,~~
2\lambda_1+3\lambda_2\leq 16\pi,\nonumber \\
&& |\lambda_5|\leq \frac{1}{2}{\rm min}\left[\sqrt{(\lambda\pm 8\pi)(\lambda_1-\lambda_2\pm 8\pi)}\right],\nonumber\\
&&|\lambda_4| \leq \frac{1}{\sqrt 2}\sqrt{\left(\lambda-\frac{8}{3}\pi\right)\left(4\lambda_1+\lambda_2-8\pi\right)}.
\label{eq:luni}
\end{eqnarray} 
\section{The Renormalization Group Equations}\label{3}
In this section, we present the RG equations (RGEs) for the scalar, gauge and Yukawa couplings relevant for our analysis. For the SM fermions, we will only keep the dominant top-quark Yukawa coupling terms. Depending on whether the renormalization scale $\mu$ is below or above the seesaw scale $M_\Delta$, the RG running will be different, as follows:
\subsection{For $\mu<M_\Delta$}
Below the seesaw scale $M_\Delta$, the heavy Higgs triplets can be integrated out to obtain a low-energy effective scalar potential for the SM Higgs doublet:
\begin{eqnarray}
V_{\rm eff}(\Phi) = -m_\Phi^2(\Phi^\dag \Phi)+\frac{1}{2}(\lambda-\lambda_6^2)(\Phi^\dag \Phi)^2,
\end{eqnarray}
and hence, the effective SM Higgs quartic coupling is shifted down: 
\begin{eqnarray}
&& \lambda \to \lambda_{\rm SM}=\lambda-\lambda_6^2\, ,\label{lamSM} \\
{\rm where}&&   
\lambda_6 = \frac{\Lambda_6}{M_\Delta}=\frac{2v_\Delta M_\Delta}{v^2}\left(1+\frac{v^2}{2M_\Delta}(\lambda_4-\lambda_5)\right)\label{eq:lam6}
\end{eqnarray} 
from Eq.~(\ref{eq:vdelta}). Thus, for a low-scale seesaw with $M_\Delta$ comparable to $v$, we always have $\lambda_6^2\ll \lambda$ below $M_\Delta$, and hence, its effect on the SM Higgs quartic coupling can be ignored. However, for a high-scale seesaw with $M_\Delta \gg v$, the effect of $\lambda_6$ could be non-negligible.  

The two-loop RG equation for the Higgs quartic coupling is given by\begin{eqnarray}
\frac{d\lambda}{d\ln\mu} = \frac{\beta_\lambda^{(1)}}{16\pi^2}+\frac{\beta_\lambda^{(2)}}{(16\pi^2)^2},
\label{lrge}
\end{eqnarray}
with the $\beta$-functions~\cite{machacek,ford,ramond,luo} 
\begin{eqnarray}
\beta_\lambda^{(1)} &=& 12\lambda^2-\left(\frac{9}{5}g_1^2+9g_2^2\right)\lambda+\frac{9}{4} \left(\frac{3}{25}g_1^4+\frac{2}{5}g_1^2g_2^2+g_2^4\right)+12y_t^2\lambda-12y_t^4,\label{self-1}\\
\beta_\lambda^{(2)} &=& -78 \lambda^3  + 18 \left( \frac{3}{5} g_1^2 + 3 g_2^2 \right) \lambda^2
 - \left( \frac{73}{8} g_2^4  - \frac{117}{20} g_1^2 g_2^2
 - \frac{1887}{200} g_1^4  \right) \lambda - 3 \lambda y_t^4
 \nonumber \\
 &&+ \frac{305}{8} g_2^6 - \frac{289}{40} g_1^2 g_2^4
 - \frac{1677}{200} g_1^4 g_2^2 - \frac{3411}{1000} g_1^6
 - 64 g_3^2 y_t^4 - \frac{16}{5} g_1^2 y_t^4
 - \frac{9}{2} g_2^4 y_t^2
 \nonumber \\
 && + 10 \lambda \left(
  \frac{17}{20} g_1^2 + \frac{9}{4} g_2^2 + 8 g_3^2 \right) y_t^2
 -\frac{3}{5} g_1^2 \left(\frac{57}{10} g_1^2 - 21 g_2^2 \right)
  y_t^2  - 72 \lambda^2 y_t^2  + 60 y_t^6.
\label{self-2}
\end{eqnarray}
The boundary condition for $\lambda(\mu)$ at a given renormalization scale 
$\mu$ can be determined from the one-loop matching condition~\cite{sirlin} for 
the SM Higgs boson pole mass $M_h$ and its running mass $m_h(\mu)=\sqrt{\lambda(\mu)}v$:~\footnote{Henceforth, we will denote the pole masses by upper case, 
and the running masses by lower case, whenever applicable.}
\beq
 \lambda(\mu) = \frac{M_h^2}{v^2} \left[ 1+ \Delta_h(\mu) \right],
\label{match-mh}
\eeq
with the expression for $\Delta_h(\mu)$ explicitly given in Appendix~\ref{A}.  
 
For the top-quark Yukawa coupling, we have the two-loop RG equation
\begin{eqnarray}
\frac{dy_t}{d\ln\mu} = \left(\frac{\beta_t^{(1)}}{16\pi^2}+\frac{\beta_t^{(2)}}{(16\pi^2)^2} \right)y_t,
\label{ytrge}
\end{eqnarray}
where~\cite{machacek} 
\begin{eqnarray}
\beta_t^{(1)} &=&\frac{9}{2} y_t^2 -
  \left(
    \frac{17}{20} g_1^2 + \frac{9}{4} g_2^2 + 8 g_3^2
  \right),  \label{topYukawa-1} \\
\beta_t^{(2)} &=&
 -12 y_t^4 +   \left(
    \frac{393}{80} g_1^2 + \frac{225}{16} g_2^2  + 36 g_3^2
   \right)  y_t^2  + \frac{1187}{600} g_1^4 - \frac{9}{20} g_1^2 g_2^2 +
  \frac{19}{15} g_1^2 g_3^2
 \nonumber\\
&&   - \frac{23}{4}  g_2^4  + 9  g_2^2 g_3^2  - 108 g_3^4+ \frac{3}{2} \lambda^2 - 6 \lambda y_t^2 .
\label{topYukawa-2}
\end{eqnarray}
The boundary condition for $y_t(\mu)$ can be determined from the matching condition between the running top quark mass $m_t(\mu) = y_t(\mu)v/\sqrt 2$ and its pole mass $M_t$,  analogous to Eq.~(\ref{match-mh}):
\begin{eqnarray}
y_t(\mu) = \frac{\sqrt 2 M_t}{v}\left[1+\Delta_t(\mu)\right],
\label{match-mt}
\end{eqnarray}
where $\Delta_t(\mu)$ gets contributions from QCD~\cite{gray, Fleischer, chety, melnikov} as well as electroweak corrections~\cite{Hempfling, kalmy}. The QCD corrections up to ${\cal O}(\alpha_3^2)$ and the electroweak corrections up to order ${\cal O}(\alpha)$ are explicitly given in 
Appendix~\ref{A}. 

The two-loop RG equations for the SM gauge couplings are given by~\cite{machacek,ramond}  
\begin{eqnarray}
\frac{dg_i}{d\ln\mu} = -\frac{g_i^3}{16\pi^2}b_i-\frac{g_i^3}{(16\pi^2)^2}\sum_{j=1}^3 b_{ij}g_j^2-\frac{g_i^3y_t^2}{(16\pi^2)^2}a_i,
\label{SMgrge}
\end{eqnarray}
where for $i=1,2,3$, the $\beta$-function coefficients are given by 
\begin{eqnarray}
b_1 &=& -\frac{2}{3} N_F - \frac{1}{10}\, , \quad 
b_2 = \frac{22}{3} -\frac{2}{3}N_F-\frac{1}{6}\, , \quad 
b_3 = 11-\frac{2}{3}N_F\, , \label{bi} \\
b_{ij} &=& \left(\begin{array}{ccc}
0 & 0 & 0 \\
0 & \frac{136}{3} & 0 \\
0 & 0 & 102
\end{array}\right) - \frac{N_F}{2}\left(\begin{array}{ccc}
\frac{19}{15} & \frac{1}{5} & \frac{11}{30} \\
\frac{3}{5} & \frac{49}{3} & \frac{3}{2}\\
\frac{44}{15} & 4 & \frac{76}{3}
\end{array}\right) - \left(\begin{array}{ccc}
\frac{9}{50} & \frac{3}{10} & 0\\
\frac{9}{10} & \frac{13}{6} & 0\\
0 & 0 & 0\end{array}\right)\, ,
\label{bij}
\end{eqnarray}
and $a_i=\left(\frac{17}{10},\frac{3}{2},2\right)$. In Eqs.~(\ref{bi}), $N_F$ is the effective number of flavors {\it below} the renormalization scale $\mu$. 
The boundary conditions for $g_i$'s are chosen to their  $\overline{\rm MS}$ values at the $Z$-pole~\cite{pdg}:  $(\alpha_1,\alpha_2,\alpha_3)(M_Z) = (0.01681, 0.03354,0.1184)$ (where $\alpha_i\equiv g_i^2/4\pi$). 

In order to simultaneously solve the coupled RGEs (\ref{lrge}), (\ref{ytrge}) and (\ref{SMgrge}), we have to impose the initial boundary conditions at a common renormalization scale. In order to do so, we first evolve the gauge coupling RGEs from $\mu=M_Z$ to $\mu=M_t$ using Eq.~(\ref{SMgrge}) without the top-Yukawa contribution and setting $N_F=5$ in Eqs.~(\ref{bi}) and (\ref{bij}). Then following Appendix~\ref{A}, we set the boundary conditions for the Higgs quartic coupling and the top-Yukawa coupling at the common scale $\mu=M_t$, and evolve them  
to $\mu<M_\Delta$ along with the gauge couplings with their full SM RGE given in Eq.~(\ref{SMgrge}) with $N_F=6$ in Eqs.~(\ref{bi}) and (\ref{bij}).   

\begin{figure}
\centering
\includegraphics[width=7cm]{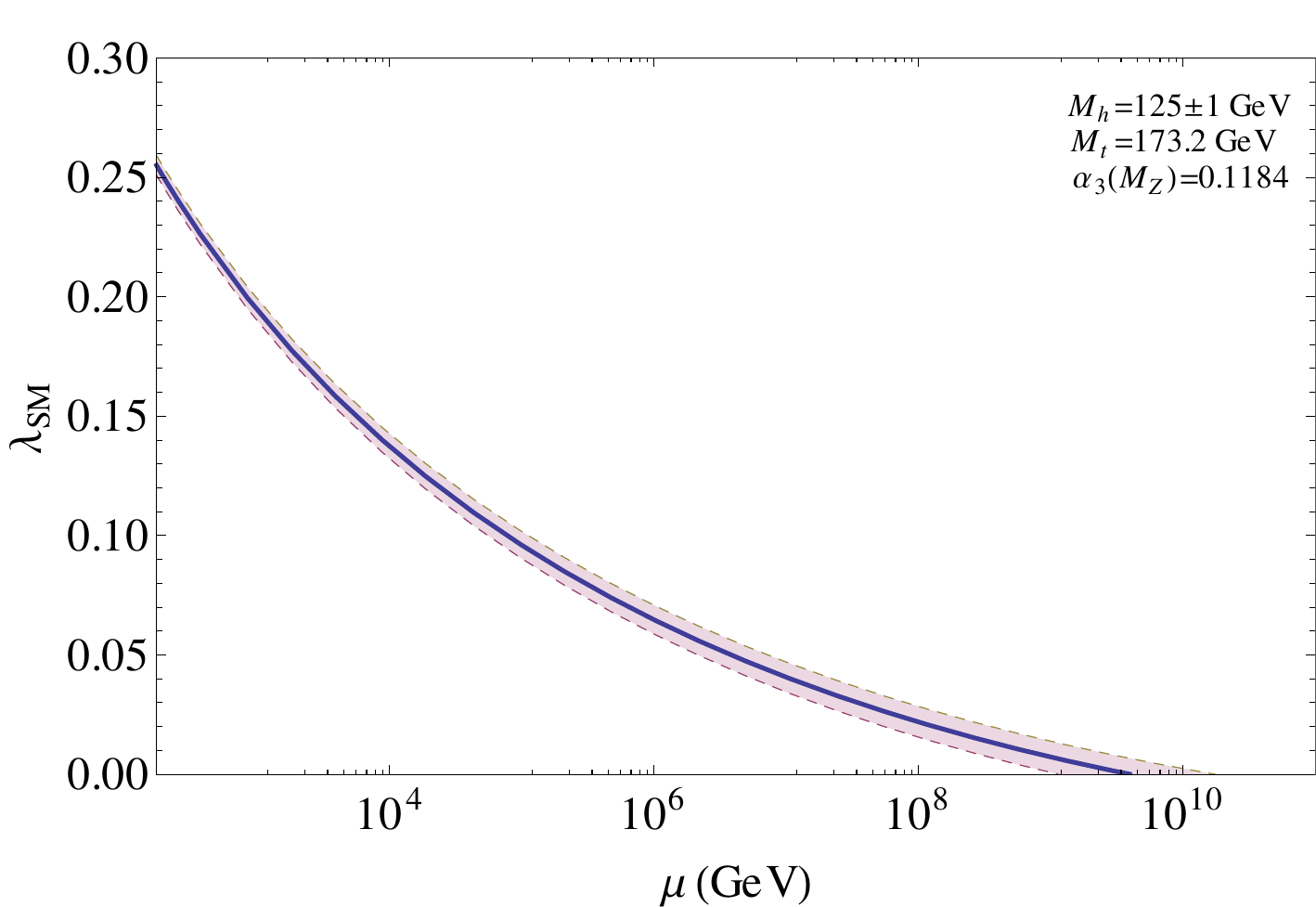}
\caption{The RG running of the SM Higgs quartic coupling for $M_h= 125\pm 1$ GeV.}
\label{fig:1}
\end{figure}

Using this procedure and for the chosen parameter values as listed in Appendix~\ref{A}, we find that the SM Higgs quartic coupling becomes negative at a renormalization scale $\mu=10^9$ - $10^{10}$ GeV for 
$M_h= 125\pm 1$ GeV, as shown in Fig.~\ref{fig:1}. Here the solid line corresponds to the running of $\lambda_{\rm SM}$ for $M_h=125$ GeV and the lower (upper) dashed line corresponds to $M_h=124~ (126)$ GeV. For $M_h=125$ GeV, we obtain 
the SM vacuum instability scale of $\Lambda_I=4\times 10^9$ GeV.   
\subsection{For $\mu\geq M_\Delta$}
For renormalization scale $\mu\geq M_\Delta$, the $\beta$-functions 
$\beta^{(1)}_t$ and $\beta^{(2)}_t$ in Eqs.~(\ref{topYukawa-1}) and
(\ref{topYukawa-2}) 
for the running of the top-quark Yukawa coupling remain unchanged since the 
coupling $ht\bar{t}$ in the type-II seesaw model is almost identical to that 
in the SM. 
However, the $\beta$-functions for the electroweak gauge couplings and the 
Higgs quartic couplings get contributions from the triplet Higgs sector. 
Following the one-loop corrections evaluated in~\cite{schmidt,chao}, and assuming that $M_\Delta>M_t$, we replace $b_i$'s in Eq.~(\ref{bi}): $b^{\rm SM}_i=\left(-\frac{41}{10},\frac{19}{6},7\right)$ with 
\begin{eqnarray}
b_i \to \left(-\frac{47}{10},\frac{5}{2},7\right)
\label{g12high}
\end{eqnarray}
and $\beta_\lambda^{(1)}$ in Eq.~(\ref{self-1}) with 
\begin{eqnarray}
\beta_\lambda^{(1)}\to \beta_\lambda^{(1)}+6\lambda_4^2+4\lambda_5^2.
\label{self-1n}
\end{eqnarray}  

For the new scalar couplings in the type-II seesaw model, the one-loop RG equations are given by~\cite{schmidt, chao}
\beq \label{Lam1}
 16\pi^2 \frac{d \lambda_1}{d \ln \mu} &=&
  -\left( \frac{36}{5} g_1^2 + 24g_2^2 \right) \lambda_1
  +\frac{108}{25}g_1^4 +18g_2^4 + \frac{72}{5}g_1^2g_2^2 \nonumber\\
  && + 14\lambda_1^2 +4 \lambda_1 \lambda_2 + 2 \lambda_2^2
  + 4 \lambda_4^2+4\lambda_5^2
  + 4 \tr\left[{\bf S}_\Delta \right] \lambda_1
  - 8 \tr\left[{\bf S}_\Delta^2 \right],  \\
\label{Lam2}
 16\pi^2 \frac{d \lambda_2}{d \ln \mu} &=&
  - \left( \frac{36}{5}g_1^2 + 24g_2^2 \right)\lambda_2
  +12g_2^4 -\frac{144}{5}g_1^2g_2^2 \nonumber\\
  && + 3 \lambda_2^2 +12 \lambda_1 \lambda_2 - 8 \lambda_5^2
  + 4 \tr \left[{\bf S}_\Delta  \right]\lambda_2
  + 8 \tr\left[{\bf S}_\Delta^2 \right],  \\
\label{Lam3}
 16\pi^2\ \frac{d \lambda_4}{d \ln \mu} &=&
 - \left( \frac{9}{2}g_1^2 + \frac{33}{2}g_2^2 \right)\lambda_4
 +\frac{27}{25}g_1^4 +6g_2^4 \nonumber\\
 && + \left( 8 \lambda_1 + 2 \lambda_2 + 6 \lambda+4\lambda_4 +6 y_t^2
 + 2\tr\left[{\bf S}_\Delta \right] \right) \lambda_4
 +8 \lambda_5^2-4\tr\left[ {\bf S}_\Delta^2 \right], \\
\label{Lam4}
 16\pi^2 \frac{d \lambda_5}{d \ln \mu} &=&
 -\frac{9}{2}g_1^2\lambda_5-\frac{33}{2}g_2^2\lambda_5
 -\frac{18}{5}g_1^2g_2^2 \nonumber\\
 &&+ \left(  2\lambda_1-2\lambda_2+2 \lambda + 8 \lambda_4 + 6 y_t^2
 + 2 \tr\left[{\bf S}_\Delta \right] \right) \lambda_5
 + 4\tr\left[{\bf S}_\Delta^2 \right] .
\eeq
Here, ${\bf S}_\Delta=Y_\Delta^\dagger Y_\Delta $
 and its corresponding RG equation is given by
\beq
 16\pi^2 \frac{d {\bf S}_\Delta}{d \ln \mu} =
 6 \; {\bf S}_\Delta^2
 -3 \left( \frac{3}{5} g_1^2 +3 g_2^2 \right) {\bf S}_\Delta
 + 2 \tr[{\bf S}_\Delta] {\bf S}_\Delta .
\label{S_Delta}
\eeq
For our numerical purposes, we will fix the structure of the Yukawa coupling 
matrix $Y_\Delta$ as in Eq.~(\ref{Ydstruct}) to fit the low-energy neutrino oscillation data, whereas the overall neutrino mass scale in Eq.~(\ref{eq:neutrino}) is fixed by the scalar triplet vev $v_\Delta$. For small $v_\Delta\sim {\cal O}({\rm eV})$, ${\rm Tr}(Y_\Delta)$ is of order unity and we cannot ignore the effect of ${\bf S}_\Delta$ on the RG equations (\ref{Lam1})-(\ref{Lam4}).    

\begin{figure}[tb]
\centering
\includegraphics[width=7cm]{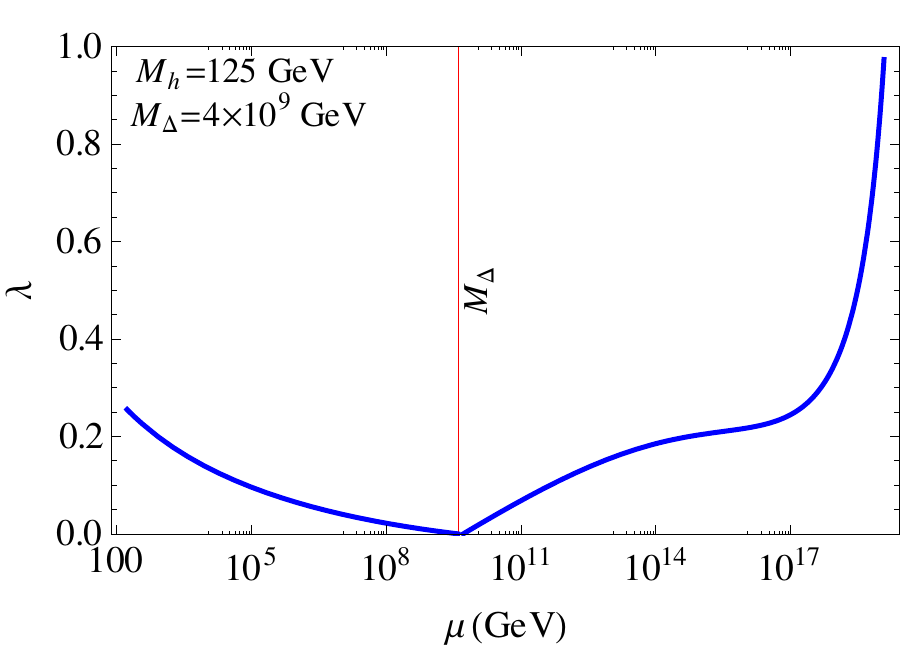}
\hspace{0.5cm}
\includegraphics[width=7cm]{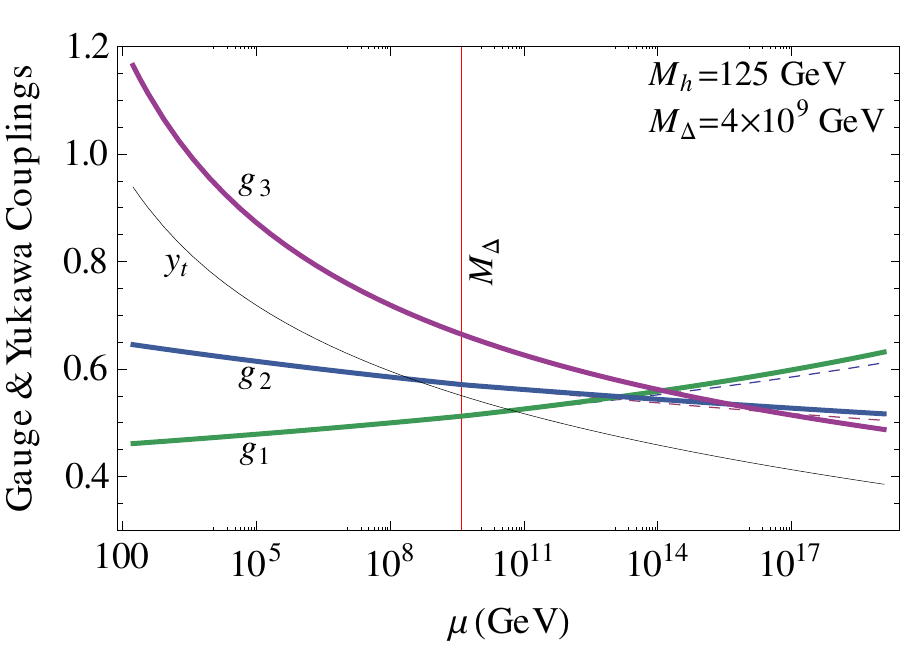}
\caption{(Left panel) The RG running of the scalar quartic coupling for $M_h=125$ GeV in the type-II seesaw model with $M_\Delta=4\times 10^{9}$ GeV. (Right Panel) Also shown are the gauge and top-Yukawa couplings. The dashed lines show the running of these parameters in the SM. The boundary conditions for the SM parameters are same as in Fig.~\ref{fig:1}.}
\label{fig:2}
\end{figure}

Note that the new contributions in Eq.~(\ref{self-1n}) are both {\it positive} which is a crucial feature in changing the overall sign of the $\beta$-function for the quartic coupling, thereby improving the electroweak vacuum stability in the type-II seesaw model. This is illustrated in Fig.~\ref{fig:2} (left panel) for $M_h=125$ GeV and the seesaw scale $M_\Delta=4\times 10^{9}$ GeV at which $\lambda_{\rm SM}$ vanishes (as already shown in Fig.~\ref{fig:1}). Here we have chosen $v_\Delta=0.05$ eV so that the effect of $\lambda_6$ on $\lambda_{\rm SM}$ in Eq.~(\ref{lamSM}) is negligible. We have shown the result for a sample set of initial values for the $\lambda_i$'s (with $i=1,2,4,5$) at $\mu=M_\Delta$: $\{0.0767, 0.0079, 0.8174, -0.3569\}$ which satisfy all the stability and unitarity conditions discussed in Section~\ref{stab}. The full allowed range of $\lambda_i$'s satisfying these conditions will be presented in the following Section.     

We have also checked that the SM gauge and Yukawa couplings remain finite up to the Planck scale for the allowed parameter space satisfying the stability and unitarity conditions in the type-II seesaw model. For illustration, we have shown the RG running of the three gauge couplings and the dominant top-Yukawa coupling in Fig:~\ref{fig:2} (right panel) for the parameter values mentioned above. The dashed lines show the RG running of the gauge couplings $g_1$ and $g_2$ in the SM which are different from those in the type-II seesaw model above the seesaw scale, as shown in Eq.~(\ref{g12high}).   
\section{The Allowed Parameter Space}\label{4}
In this section, we analyze the allowed parameter space in the scalar sector of the type-II seesaw model which yields a doublet Higgs pole mass $M_h=125$ GeV while satisfying all the vacuum stability and unitarity conditions discussed in Section~\ref{stab} up to the Planck scale and also satisfying the neutrino oscillation data at low-scale. We present our scan results for two benchmark scenarios as follows:
\subsection{Low-scale Seesaw}\label{4a}
In our first benchmark scenario, we would like to consider a low-scale type-II seesaw model which could be testable at the LHC and other low-energy experiments~\cite{han,goran} (for an earlier review, see e.g.,~\cite{acco}). First, let us review the existing constraints on the seesaw scale. The strongest limits come from the ongoing LHC searches for doubly-charged Higgs bosons~\cite{atlas-typeII,cms-typeII}. They can be produced via $q\bar q\to \gamma^*,Z^*,W^{\pm *}W^{\pm *}\to H^{++}H^{--},~q'\bar{q}\to W^*\to H^{\pm\pm}H^\mp,H^{\pm\pm}W^\mp$ and have the following possible decay channels: (i) same-sign charged lepton pair ($\ell^\pm\ell^\pm$), (ii) pair of charged gauge bosons ($W^\pm W^\pm$), (iii) $W^\pm H^\pm$, and (iv) $H^\pm H^\pm$, if kinematically allowed. For $v_\Delta<10^{-4}$ GeV (large Yukawa couplings) and degenerate triplet scalars, the doubly-charged Higgs decays dominantly to $\ell^\pm\ell^\pm$, and the current 95\% CL lower limit on its mass is $M_{H^{\pm\pm}}>$ 300 - 400 GeV~\cite{atlas-typeII}, depending on the final-state lepton-flavor. On the other hand, for $v_\Delta>10^{-4}$ GeV (small Yukawa), the branching ratio to $\ell^\pm\ell^\pm$ decreases significantly, and the other decay channels (ii), (iii) and (iv) become dominant. In this case, the lower limit on the mass of $H^{\pm\pm}$ can be lowered to about 100 GeV, 
provided the mass splitting between the singly- and doubly-charged scalars is 
large enough to allow for the cascade decays~\cite{goran}.  Note that in either case, the constraints from other low-energy experiments such as the lepton-flavor violating decays can be satisfied, provided~\cite{Akeroyd:2009nu, Tsumura} 
\begin{eqnarray}
     v_\Delta M_{H^{\pm\pm}} \gsim 150~{\rm eV~GeV}\, . 
\label{lfv}
\end{eqnarray}

For the singly-charged Higgs bosons $H^\pm$, the coupling to a pair of quarks is suppressed by $v_\Delta/v$ in the type-II seesaw model. Hence, the conventional mechanisms for its production at hadron colliders such as $gg\to tbH^+$ and $bg\to tH^+$ are suppressed. Moreover, the branching ratio of $t\to bH^+$ would also be suppressed, and the LHC limits on $M_{H^\pm}$~\cite{atlas-hp,cms-hp} may not apply in this case~\footnote{Due to similar suppression in their production and decay rates, the dominantly triplet neutral scalars $H^0$ and $A^0$ can also evade the mass limits from the direct LHC searches for a neutral Higgs boson~\cite{atlas-neutral,cms-neutral}.}. However, we can still apply the combined LEP lower limit on $M_{H^\pm}$ of about 80 GeV~\cite{lep}. 

Apart from these constraints from direct searches, the triplet Higgs sector also contributes to the electroweak precision observables, namely, the $S,T,U$ parameters~\cite{lavoura}.  The dominant constraint comes from the $T$ parameter 
which is governed by the mass difference between the singly- and doubly-charged Higgs bosons, $\Delta M\equiv  |M_{H^{\pm\pm}}-M_{H^\pm}|$. For a light SM 
Higgs, it was shown in Ref.~\cite{goran} that the allowed range of $\Delta M$ is 
roughly 0 - 50 GeV, and using the latest best-fit results for the oblique 
parameters, Ref.~\cite{sharma} updated it to $\Delta M\lsim 40$ GeV, almost independently of the doubly-charged Higgs mass.  

\begin{figure}
\vspace{-1.5cm}
\centering
\begin{tabular}{c c}
\includegraphics[width=7cm]{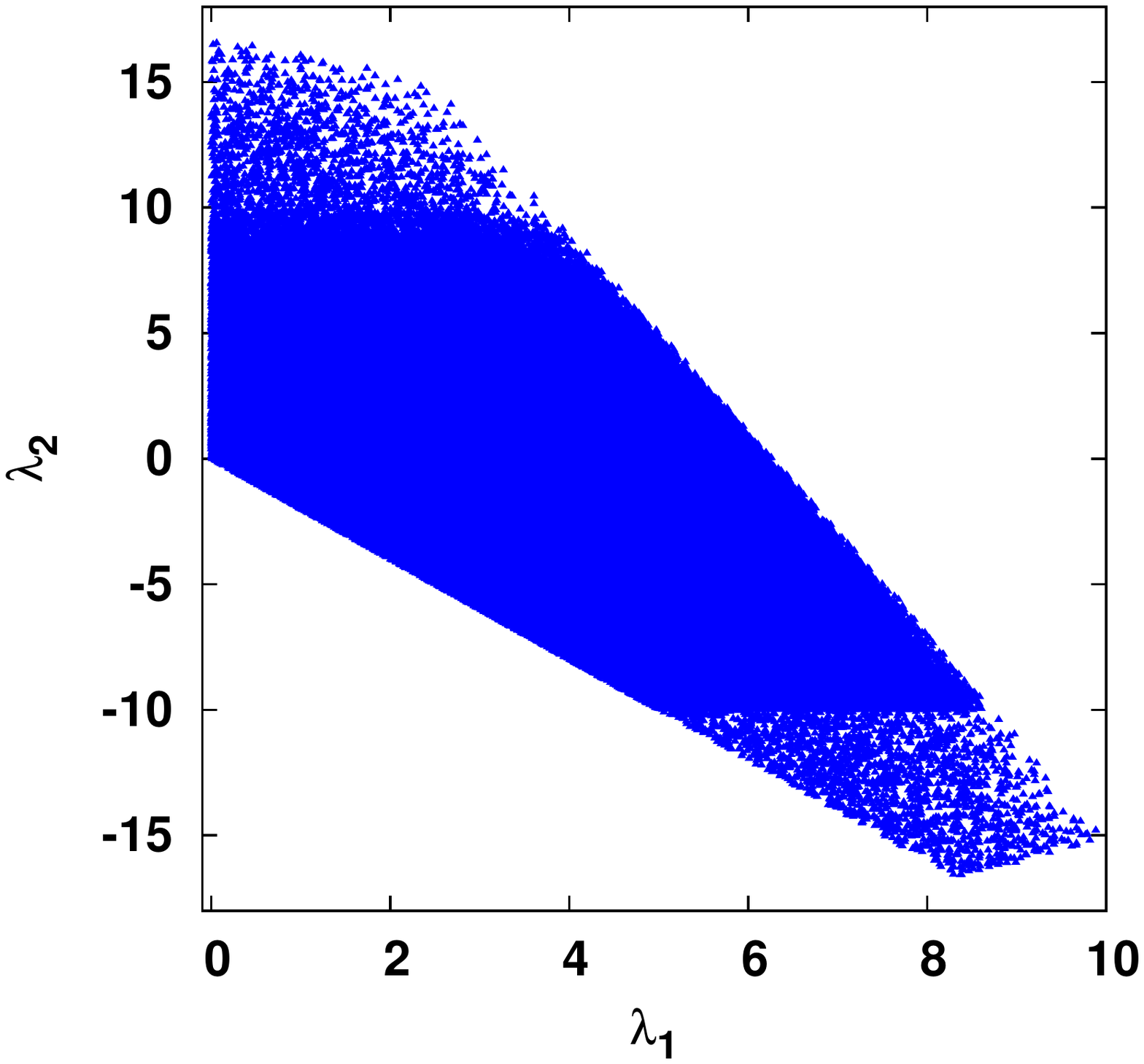}\hspace{0.5cm} &
\includegraphics[width=7cm]{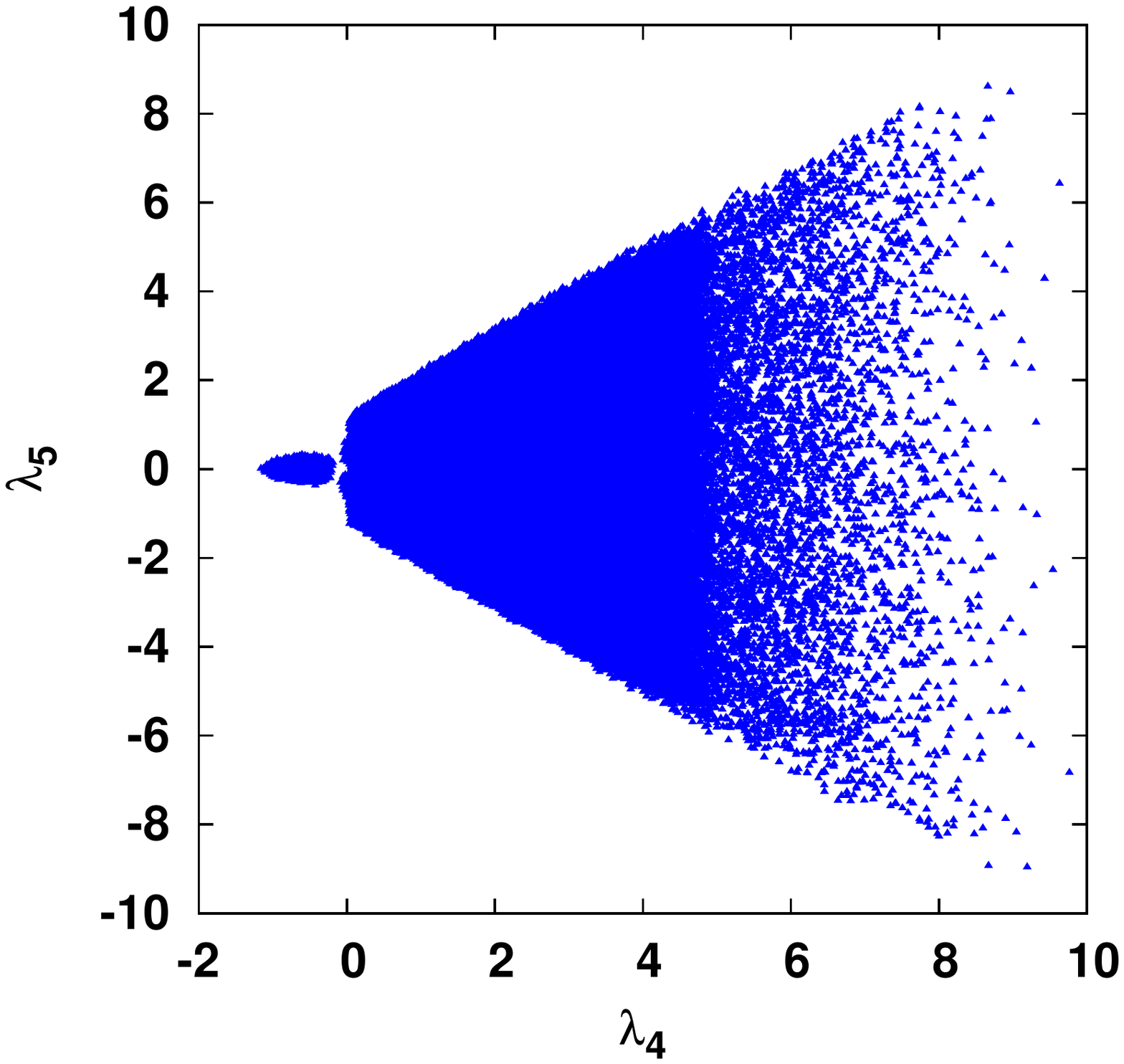} \\
& \\ [-2.5cm]
(a) 
& 
\hspace{0.5cm} 
(b)
\end{tabular}
\caption{The allowed parameter space in (a) ($\lambda_1,\lambda_2$) plane and (b) $(\lambda_4, \lambda_5)$ plane for the scalar sector of low-scale ($M_\Delta = 200$ GeV,$v_\Delta = 1$ GeV) type-II seesaw model.}
\label{fig:low}
\end{figure}

In view of these constraints on the scalar sector, we choose a seesaw scale of $M_\Delta=200$ GeV in our first benchmark scenario. For $M_h=125$ GeV, this requires the scalar quartic coupling at $M_\Delta$ to be $\lambda(\mu=200~{\rm GeV})=0.25$ (cf. Fig.~\ref{fig:1}). With this initial value of $\lambda$, we perform scans over the parameter space of the remaining scalar couplings to obtain the allowed range satisfying the vacuum stability and perturbative conditions discussed in Section~\ref{stab}. Our results are presented in Fig.~\ref{fig:low}. 
Here we choose $v_\Delta=1$ GeV for illustration, but our scan 
results are independent of the exact value of $v_\Delta$ in the limit 
$v_\Delta\ll v$ as long as 
$v_\Delta\gsim 10^{-4}$ GeV in order to be able to avoid the collider 
constraints on $M_{H^{\pm\pm}}$ for a low seesaw scale, as discussed above. 

Note that the region around $(\lambda_4,\lambda_5)=(0,0)$ is not allowed since the RGE for the scalar quartic coupling in 
the vicinity of this region is almost identical to its SM RGE [cf. Eq.~(\ref{self-1n})], and hence, we hit the SM vacuum instability scale below $M_P$. Nonetheless, we find that there exists a large parameter space in a low-scale type-II seesaw model which yield a SM Higgs mass of 125 GeV while satisfying all the stability and unitarity constraints as well as the current experimental bounds. Some phenomenological implications of this result will be discussed in Section~\ref{5}. 
\subsection{High-scale Seesaw}\label{4b}
\begin{figure}
\vspace{-1.5cm}
\centering
\begin{tabular}{c c}
\includegraphics[width=7cm]{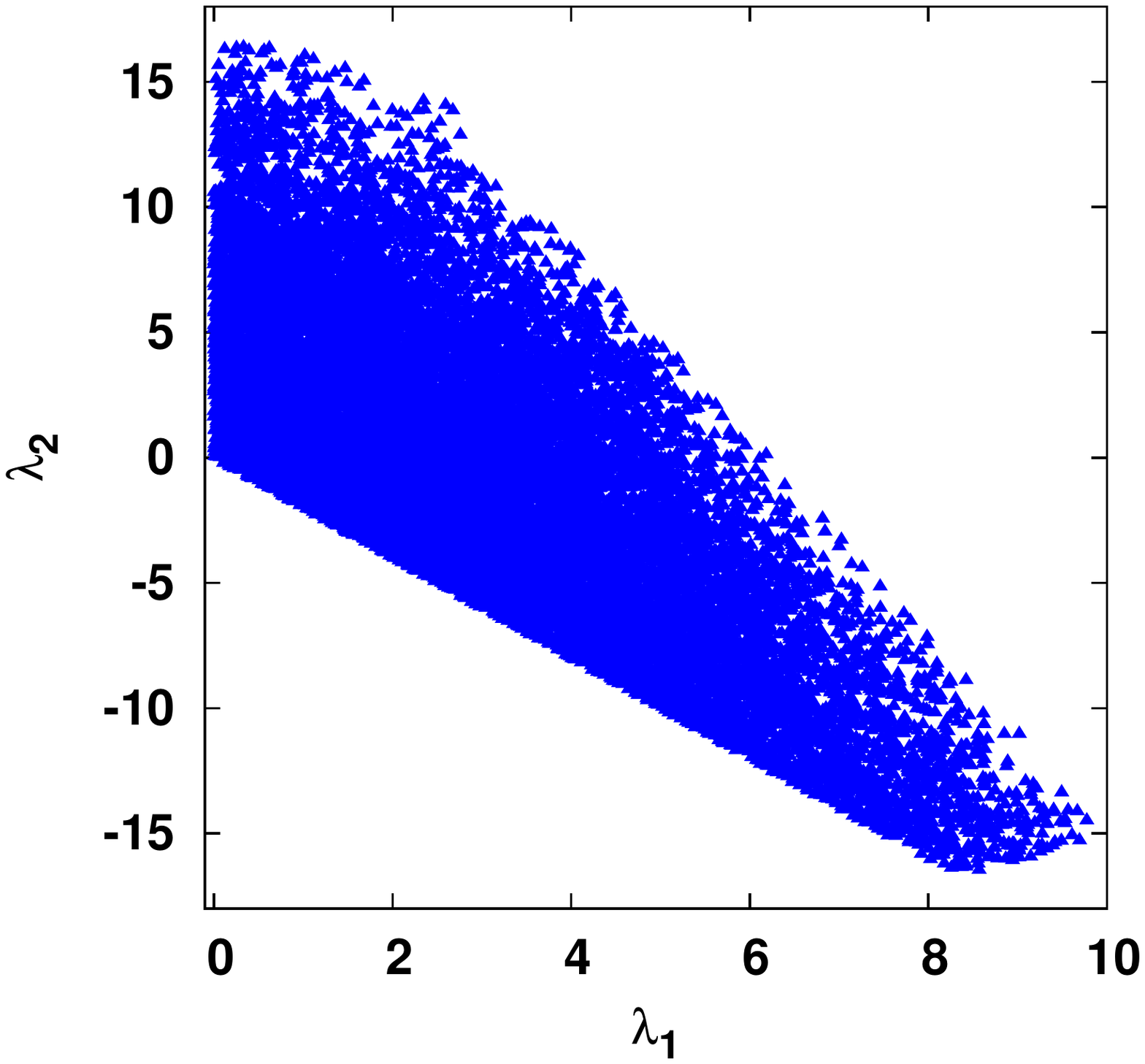}\hspace{0.5cm} &
\includegraphics[width=7cm]{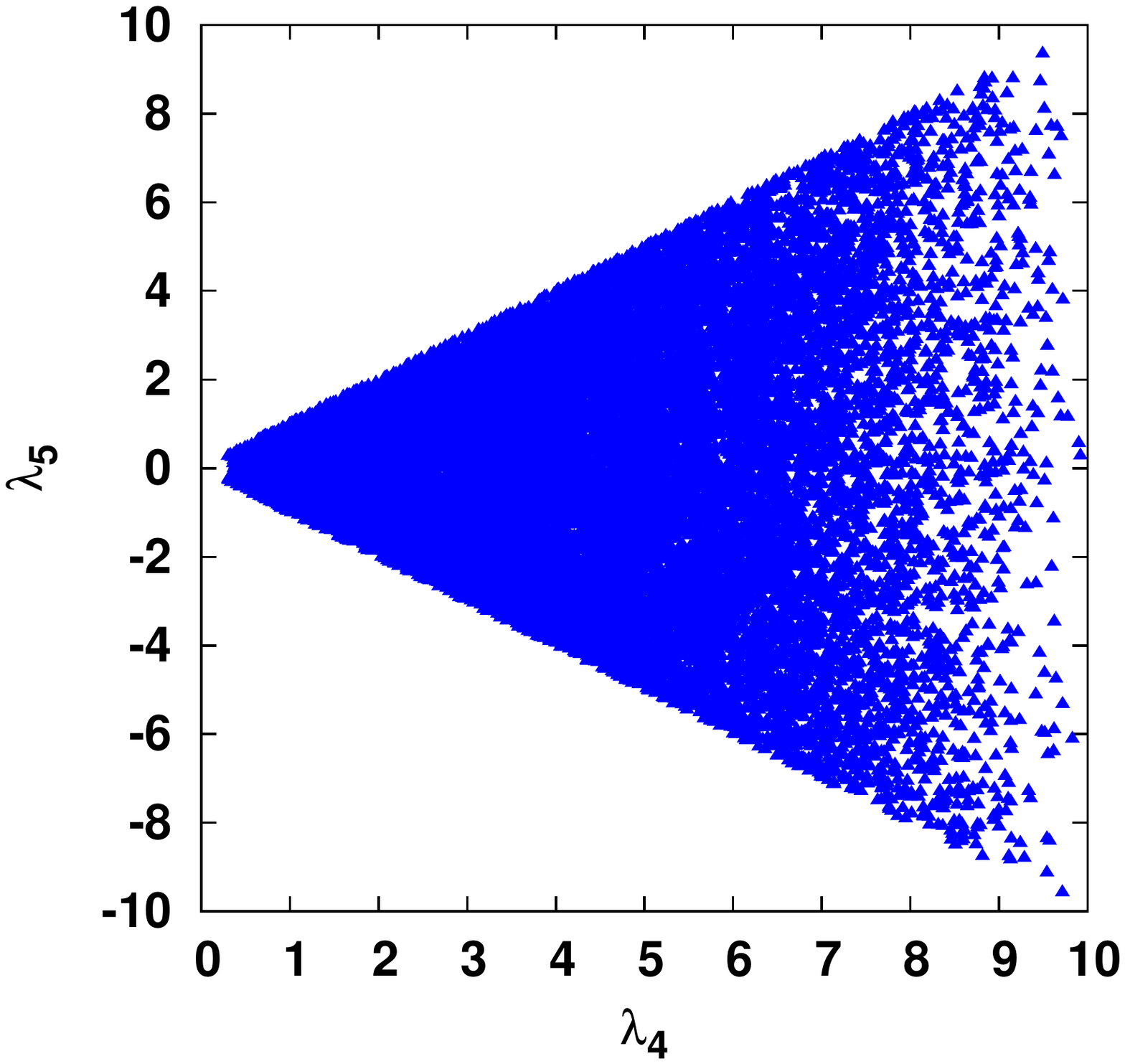} \\
& \\ [-2.5cm]
(a) & 
\hspace{0.5cm} (b)
\end{tabular}
\caption{The allowed parameter space in (a) ($\lambda_1,\lambda_2$) plane and (b) $(\lambda_4, \lambda_5)$ plane for the scalar sector of a high-scale ($M_\Delta = 4\times10^{9}$ GeV,$v_\Delta = 0.05$ eV) type-II seesaw model.}\label{fig:high}
\end{figure}

From the point of view of vacuum stability, the highest possible seesaw scale is determined by the vanishing SM Higgs quartic coupling at $\mu=\Lambda_I$. For $M_h=125$ GeV, $\lambda_{\rm SM}$ becomes zero at $\Lambda_I = 4\times 10^{9}$ GeV (Fig.~\ref{fig:1}). Hence, in our second benchmark scenario, we fix the seesaw scale to be $M_\Delta=\Lambda_I=4\times 10^9$ GeV. For such a high seesaw scale, the triplet scalar masses are well beyond the current experimental constraints~\cite{atlas-typeII,cms-typeII}. Hence, the triplet vev $v_\Delta$ can be anywhere between its upper bound of ${\cal O}(1)$ GeV and ${\cal O}(0.01)$ eV below which the Yukawa couplings in Eq.~(\ref{YD}) become non-perturbative. For our analysis of the parameter space in this benchmark scenario, we have chosen $v_\Delta=0.05$ eV so that $\lambda_6$ in Eq.~(\ref{eq:lam6}) is small and its effect on $\lambda$ can be neglected~\footnote{For a non-negligible value of $\lambda_6$, the quartic coupling for $M_h=125$ GeV will become negative at a {\it lower} value of $\mu$, and we can do a similar analysis by fixing the seesaw scale to this value.}. Using the initial condition $\lambda=0$ at $\mu=M_\Delta$, we scan over the rest of the scalar couplings at this scale to find the allowed parameter space satisfying the stability and perturbativity conditions given by Eqs.~(\ref{eq:lstab}) and (\ref{eq:luni}). Our results are shown in Fig.~\ref{fig:high}. We find that the allowed parameter space is almost identical to the low-scale seesaw case (Fig.~\ref{fig:low}), except for the {\it negative} values of $\lambda_4$ which are not allowed in the high-scale seesaw scenario.   

Even though such a high-scale seesaw model is not accessible at colliders, it 
could be useful in explaining the observed matter-antimatter asymmetry in our Universe via leptogenesis~\cite{Ma:1998}~\footnote{We could also have 
leptogenesis with a low-scale type-II seesaw model through a resonant 
mechanism, but it requires more than one scalar triplets. For a review of leptogenesis in both high- and low-scale type-II seesaw models, see e.g.,~\cite{Hambye:2012}.} as well as inflation~\cite{sahu,Chen:2010} through the decay of the heavy scalar triplet. Moreover, 
the minimal type-II seesaw model we have been discussing here can be easily  
extended to accommodate a Dark Matter candidate by adding  
extra SM singlet scalar field(s) with a discrete $Z_2$ symmetry~\cite{musolf, Kanemura:2012rj, Wang:2012ts}. A detailed phenomenological analysis of 
these possibilities in the parameter space allowed by vacuum stability and perturbativity as illustrated here for the minimal model is left for future study.   
\section{Predictions for the Decay Rates of $h\to \gamma \gamma,~Z\gamma$ }
\label{5}
In the SM, the decay $h\to \gamma \gamma (Z)$ is mediated at the one-loop 
level by the virtual exchange of SM fermions (dominantly the top-quark) and 
the $W$-boson~\cite{hgg,shifman}. The Higgs-to-diphoton decay channel is one of the highest mass resolution channels and plays 
an important role in the SM Higgs discovery at the LHC~\cite{atlas, cms}. 
The current signal strength (as defined by the ratio of the observed cross-section times branching ratio over the SM expected value) 
in the $ pp \to h \to \gamma \gamma $ channel 
is $1.8\pm 0.4$ (ATLAS)~\cite{atlas-gg} and 
$1.56\pm 0.43 $ (CMS)~\cite{cms-gg}~\footnote{The updated ATLAS analysis gives $1.65^{+0.34}_{-0.30}$~\cite{ATLASnew}, consistent with their earlier result, whereas the updated CMS analysis gives a much lower value of $0.78\pm 0.27$~\cite{CMSnew}. In the absence of a consensus between the two results, we will use the ATLAS value as our reference point in the following analysis.}. Thus there is about $2\sigma$ discrepancy 
between the observed value and the SM prediction in the $\gamma\gamma$ channel at the moment~\footnote{The theoretical uncertainties of about 30\% in the cross-section of the SM Higgs boson production through gluon fusion, $\sigma(gg\to h)$, can reduce this discrepancy between the measured and SM expected values to about $1\sigma$ level~\cite{rohini}.}. If this excess still 
remains even with higher statistics, it could be an indication for some low-scale beyond SM physics where new charged particles that couple to the SM Higgs boson
contribute constructively to the $h \to \gamma \gamma $ amplitude. 
In the type-II seesaw model, there are 
additional contributions from the new charged Higgs states~\cite{arhrib1}. 
Following the general results for spin-0, spin-1/2 and spin-1 contributions 
to the $h\to \gamma\gamma$ rate~\cite{shifman} (see also~\cite{review,spira,gunion}), 
we obtain for its partial decay width:
\begin{eqnarray}
\label{eq:THM-h2gaga}
\Gamma(h \rightarrow\gamma\gamma)
& = & \frac{\alpha^2 G_F M_{h}^3}
{128\sqrt{2}\pi^3} \bigg| \sum_f N_c Q_f^2 g_{h f\bar{f}} 
A^h_{1/2}
(\tau_f) + g_{h W^+W^-} A^h_1 (\tau_W) \nonumber \\
&& \hspace{1.5cm} + \tilde{g}_{h H^\pm\,H^\mp}
A^h_0(\tau_{H^{\pm}})+
 4 \tilde{g}_{h H^{\pm\pm}H^{\mp\mp}}
A^h_0(\tau_{H^{\pm\pm}}) \bigg|^2 \, .
\label{partial_width_htm}
\end{eqnarray}
Here $G_F$ is the Fermi coupling constant, $\alpha$ is the fine-structure 
constant, $N_c=3 (1)$ for quarks (leptons), $Q_f$ is the electric charge of 
the fermion in the loop, and $\tau_i=M_h^2/4M_i^2~(i=f,W,H^\pm,H^{\pm\pm})$. 
The first two terms in the squared amplitude (\ref{partial_width_htm}) are the 
SM fermion and $W$-boson contributions respectively, whereas the last two terms
 correspond to the $H^\pm$ and $H^{\pm\pm}$ contributions. The relevant loop 
functions are given by 
\begin{eqnarray}
A_{0}(\tau) &=& -[\tau -f(\tau)] \tau^{-2} \, ,
\label{eq:Ascalar}\\
A_{1/2}(\tau)&=& 2\left[\tau+(\tau-1)f(\tau)\right]\tau^{-2}, 
\label{eq:Afermion}\\
A_1(\tau)&=& -\left[2\tau^2+3\tau+3(2\tau-1)f(\tau)\right]\tau^{-2}, 
\label{eq:Avector} 
\end{eqnarray}
 and the function $f(\tau)$ is given by
\begin{eqnarray}
f(\tau)=\left\{
\begin{array}{ll}  \displaystyle
\left[\sin^{-1}\left(\sqrt{\tau}\right)\right]^2, & (\tau\leq 1) \\
\displaystyle -\frac{1}{4}\left[ \log\left(\frac{1+\sqrt{1-\tau^{-1}}}
{1-\sqrt{1-\tau^{-1}}}\right)-i\pi \right]^2, \hspace{0.5cm} & (\tau>1) \, .
\end{array} \right. 
\label{eq:ftau} 
\end{eqnarray}
Note that the $H^{\pm\pm}$ contribution in the amplitude of 
Eq.~(\ref{partial_width_htm}) is enhanced by a factor of four compared to 
the $H^\pm$ contribution since $H^{\pm\pm}$ has an electric charge of 
$\pm 2$ units. 

The couplings of $h$ to the SM fermions and vector bosons {\it relative} to 
the SM Higgs couplings are given by
\begin{eqnarray}
g_{hf\bar f} = \frac{\cos\alpha}{\cos\beta'}\, ,\quad
g_{hW^+W^-} = \cos\alpha+2\sin\alpha\frac{v_\Delta}{v}.
\label{coupSM}
\end{eqnarray}
From Eqs.~(\ref{mix1}) and (\ref{mix3}), we see that in the limit 
$v_\Delta\ll v$, we have $\cos\beta',\cos\alpha \sim 1$, and hence, the 
couplings in Eq.~(\ref{coupSM}) are essentially identical to the SM Higgs boson
 couplings to fermions and vector bosons. For the scalar trilinear couplings 
in Eq.~(\ref{partial_width_htm}), we have   
\begin{eqnarray}
\tilde{g}_{h H^{++}H^{--}}  & = &  \frac{M_W}{ g M_{H^{\pm \pm}}^2} g_{h H^{++}H^{--}} \, , \quad
\tilde{g}_{h H^+H^-} = \frac{M_W}{g M_{H^{\pm}}^2} g_{h H^+H^-} \, ,
\label{eq:redgcallittlehHp}
\end{eqnarray}
with the following definitions in terms of the parameters of the scalar 
potential (up to ${\cal O}(v_\Delta^2)$)~\cite{arhrib0}:
\begin{eqnarray}
g_{hH^{++}H^{--}} &=& (\lambda_1+\lambda_2)v_\Delta \sin\alpha 
+ (\lambda_4+\lambda_5)v\cos\alpha \, , \label{gh1}\\
g_{hH^{+}H^{-}}  &=& \left[\left(\lambda_1 \cos^2\beta'+(\lambda_4+\lambda_5)
\sin^2\beta'\right)v_\Delta+\sqrt 2 \lambda_5 \cos\beta'\sin\beta' v\right]
\sin\alpha + \nonumber\\
&& \left[\left(\lambda \sin^2\beta'+\lambda_4 \cos^2\beta'\right)v+\sqrt 2\cos\beta'\sin\beta' \left(\frac{2M_\Delta^2}{v^2}+\lambda_4\right)v_\Delta\right]\cos\alpha \label{gh2}
\end{eqnarray}
In the limit $v_\Delta\ll v$, we can rewrite Eqs.~(\ref{gh1}) and (\ref{gh2}) 
in the following simple forms:
\begin{eqnarray}
 g_{hH^{++}H^{--}} \simeq (\lambda_4+\lambda_5)v \, , \quad g_{hH^{+}H^{-}}\simeq \lambda_4 v. \label{ghsimp}
\end{eqnarray} 
Thus, the signs of the couplings $g_{hH^{++}H^{--}}$ and $g_{hH^{+}H^{-}}$, 
and hence, those of the $H^\pm$ and $H^{\pm\pm}$ contributions to the amplitude
in Eq.~(\ref{partial_width_htm}) are respectively fixed by the scalar 
couplings $(\lambda_4+\lambda_5)$ and $\lambda_4$ which are in turn constrained by the vacuum stability and unitarity conditions, as shown in the previous section. Due to the enhancement factor of four for the $H^{\pm\pm}$ contribution in Eq.~(\ref{partial_width_htm}), we would expect this term to dominate over the $H^\pm$ contribution for most of the allowed parameter space.   

For the partial decay width of $h \to Z \gamma $, we obtain~\cite{hzg1, hzg2, 
carena, frank}
\begin{eqnarray}
\Gamma(h \to Z \gamma) &=& 
\frac{\alpha G_F^2 M_W^2M_h^3}{64\pi^4}\left( 1 -\frac{M_Z^2}{M_h^2} \right)^3\bigg|
\frac{1}{c_W}\sum_f N_cQ_f(2I_3^f-4Q_fs_W^2)g_{hf\bar{f}}A^h_{1/2}(\tau_h^f,\tau_Z^f)\nonumber\\
&&+c_W g_{hW^+W^-}A_1^h(\tau_h^W,\tau_Z^W)
-2s_Wg_{ZH^\pm H^\mp}\tilde{g}_{hH^\pm H^\mp}A^h_0(\tau_h^{H^\pm},\tau_Z^{H^\pm})\nonumber\\
&&-4s_Wg_{ZH^{\pm\pm}H^{\mp\mp}}\tilde{g}_{hH^{\pm\pm}H^{\mp\mp}}A_0^h(\tau_h^{H^{\pm\pm}},\tau_Z^{H^{\pm\pm}}) 
\bigg|^2, 
\label{eq:hzg}
\end{eqnarray}
where $\tau_h^i=4M_i^2/M_h^2,$ $\tau_Z^i=4M_i^2/M_Z^2$ (with $i=f,W,H^\pm,H^{\pm\pm}$), and the loop-factors are given by 
\begin{eqnarray}
A^h_0(\tau_h,\tau_Z) &=& I_1(\tau_h,\tau_Z),\nonumber \\
A^h_{1/2}(\tau_h,\tau_Z) &=& I_1(\tau_h,\tau_Z)-I_2(\tau_h,\tau_Z),\\
A^h_1(\tau_h,\tau_Z) &=& 4(3-\tan^2\theta_W)I_2(\tau_h,\tau_Z)+\left[(1+2\tau_h^{-1})\tan^2\theta_W-(5+2\tau_h^{-1})\right]I_1(\tau_h,\tau_Z).\nonumber
\end{eqnarray} 
The functions $I_1$ and $I_2$ are given by 
\begin{eqnarray}
I_1(\tau_h,\tau_Z) &=&  \frac{\tau_h \tau_Z}{2\left(\tau_h-\tau_Z\right)}
+\frac{\tau_h^2\tau_Z^2}{2\left(\tau_h-\tau_Z\right)^2}
\left[f\left(\tau_h^{-1}\right)-f\left(\tau_Z^{-1}\right)\right]
+\frac{\tau_h^2\tau_Z}{\left(\tau_h-\tau_Z\right)^2}
\left[g\left(\tau_h^{-1}\right)-g\left(\tau_Z^{-1}\right)\right],\nonumber\\
I_2(\tau_h,\tau_Z) &=& -\frac{\tau_h\tau_Z}{2(\tau_h-\tau_Z)}\left[f\left(\tau_h^{-1}\right)-f\left(\tau_Z^{-1}\right)\right],
\end{eqnarray}
where the function $f(\tau)$ is defined in Eq.~(\ref{eq:ftau}), and the function $g(\tau)$ is defined as
\begin{eqnarray}
g(\tau)=\left\{
\begin{array}{ll}  \displaystyle
\sqrt{\tau^{-1}-1}\sin^{-1}\left(\sqrt{\tau}\right), & (\tau< 1) \\
\displaystyle 
\frac{1}{2}\sqrt{1-\tau^{-1}}\left[ \log\left(\frac{1+\sqrt{1-\tau^{-1}}}
{1-\sqrt{1-\tau^{-1}}}\right)-i\pi\right], \hspace{0.5cm} & (\tau\geq 1) \, .
\end{array} \right. 
\label{eq:gtau} 
\end{eqnarray}
The scalar couplings $g_{hf\bar{f}}$ and $g_{hW^+W^-}$ are given in Eq.~(\ref{coupSM}), and the scalar trilinear couplings 
$\tilde{g}_{h H^{\pm}H^{\mp}}$ and $\tilde{g}_{h H^{\pm\pm}H^{\mp\mp}}$ are given in Eq.~(\ref{eq:redgcallittlehHp}). The 
remaining couplings in Eq.~(\ref{eq:hzg}) are given by 
\begin{eqnarray}
g_{Z H^{+} H^{-}} = - \tan \theta_W\, , \quad  g_{Z H^{++} H^{--}} = 2\cot 2\theta_W \, .
\end{eqnarray}

In the SM, the partial decay width of $h\to \gamma\gamma$ is dominated by the 
$W$-loop contribution which interferes destructively with the sub-dominant 
top-loop contribution~\cite{spira}. Hence, in the type-II seesaw model, 
we can have enhancement in the $h\to \gamma\gamma$ decay width with respect 
to the SM value provided we have constructive interference of the 
charged-Higgs contributions in Eq.~(\ref{partial_width_htm}) with the 
$W$-loop contribution. This happens for $(\lambda_4+\lambda_5)<0$ which is 
allowed over a small range of the parameter space as shown in 
Fig.~\ref{fig:low} (b). 

\begin{figure}
\vspace{-1.5cm}
\centering
\begin{tabular}{c c}
\includegraphics[width=7cm]{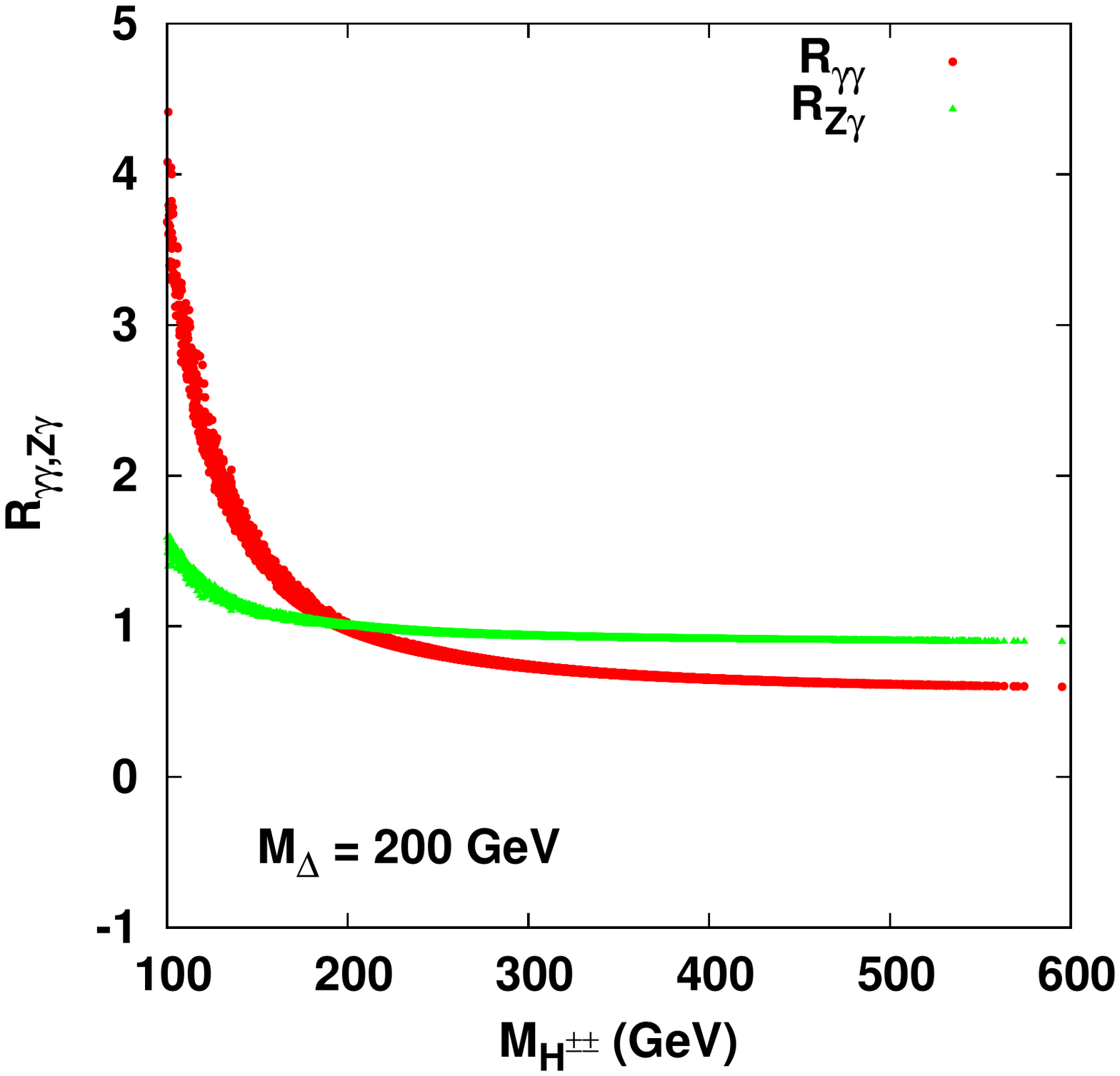}\hspace{0.5cm} & 
\includegraphics[width=7cm]{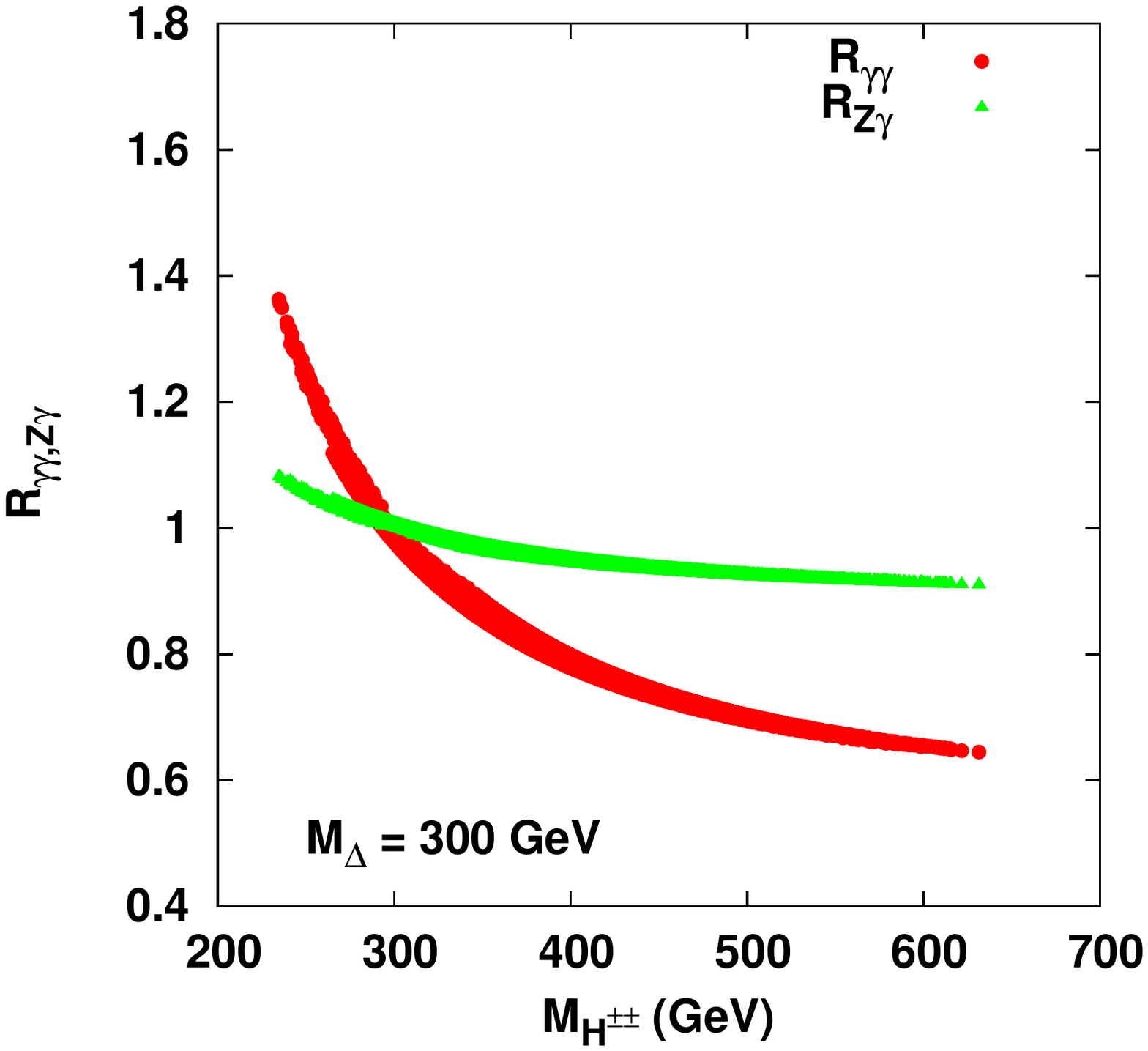} \\
& \\ [-2.5cm]
(a) & \hspace{0.5cm} (b)
\end{tabular}
\caption{The predictions for $R_{\gamma\gamma}$ (red/dark) and 
$R_{Z\gamma}$ (green/light) in the allowed parameter space of  a low-scale 
type-II seesaw model with (a) $M_\Delta = 200$ GeV and 
(b) $M_\Delta = 300$ GeV. Here we have chosen $v_\Delta=1$ GeV. 
}\label{fig:Rgg1}
\end{figure}

The doubly-charged scalar contribution dominates over the singly-charged scalar
contribution for both $h\to \gamma\gamma$ and $h\to Z\gamma$ amplitude, and 
with the same sign with respect to the SM contribution. Thus, for $(\lambda_4+\lambda_5)<0$, the decay-width is
enhanced for both $h\to \gamma\gamma$ and $h\to Z\gamma$.
For the same reason, the region where both $\lambda_4$ and $\lambda_4 + \lambda_5$ are positive, the behavior reverses. In other words, the $h\to \gamma\gamma$ and $h\to Z\gamma$ partial decay widths are correlated which can be easily seen from Fig.~\ref{fig:Rgg1} (discussed below).

In order to compare the model predictions for the signal strength with the SM value at the LHC, the partial decay widths of the processes $h\to \gamma\gamma,~Z\gamma $ 
can be expressed in terms of the following simple ratios:
\begin{eqnarray}
R_{\gamma\gamma}&=&\frac{\sigma_{\rm model}(pp\to h \to \gamma\gamma)}{\sigma_{\rm SM}(pp\to h \to \gamma\gamma)}
= \frac{\sigma_{\rm model}(pp\to h)}{\sigma_{\rm SM}(pp\to h)}\frac{{\rm BR}_{\rm model}(h\to \gamma\gamma)}{{\rm BR}_{\rm SM}(h\to \gamma\gamma)}\, ,
\label{rggdef}
\end{eqnarray}
and similarly for $R_{Z\gamma}$. 
For the dominant Higgs production channel at the LHC, namely $gg\to h$ (see e.g.,~\cite{Dittmaier}), the ratio of the production cross-sections in Eq.~(\ref{rggdef}),  
	$\sigma_{\rm model}(pp\to h)/\sigma_{\rm SM}(pp\to h) = \cos^2\alpha $
for the type-II seesaw model, where the mixing angle $\alpha$ is given by Eq.~(\ref{mix3}). Thus, for a SM-like Higgs regime of the type-II seesaw model, the 
dominant production cross-section is essentially the same as that in the SM. 
The branching ratios of all the Higgs decay channels are also the same as in the SM, except for $\gamma\gamma$ and $Z\gamma$ channels which can differ significantly as discussed above, but their contribution to the total decay width remains negligible as in the SM. Hence, for our numerical purposes, we can simply 
assume $R_{\gamma\gamma}$ defined in Eq.~(\ref{rggdef}) to be the ratio of the partial decay widths for $h\to \gamma\gamma$ in the type-II seesaw model and in the SM. 

The predictions for the ratios $R_{\gamma\gamma,~Z\gamma}$ in the allowed model parameter space for a low seesaw scale of $M_\Delta = 200$ GeV (cf. Fig.~\ref{fig:low}) are shown in Fig.~\ref{fig:Rgg1}(a), as a function of the doubly-charged Higgs mass given by Eq.~(\ref{doub-mass}) evaluated at the seesaw scale $\mu=M_\Delta$. The results for another seesaw scale $M_\Delta = 300$ GeV are shown in Fig.~\ref{fig:Rgg1}(b). We can clearly see the correlation between the $\gamma\gamma$ and $Z\gamma$ rates, as argued above. We find that the size of the enhancement in the $Z\gamma$ channel is much smaller compared to that in the $\gamma\gamma$ channel. 

\begin{figure}[h!]
\vspace{-1.5cm}
\centering
\includegraphics[width=7cm]{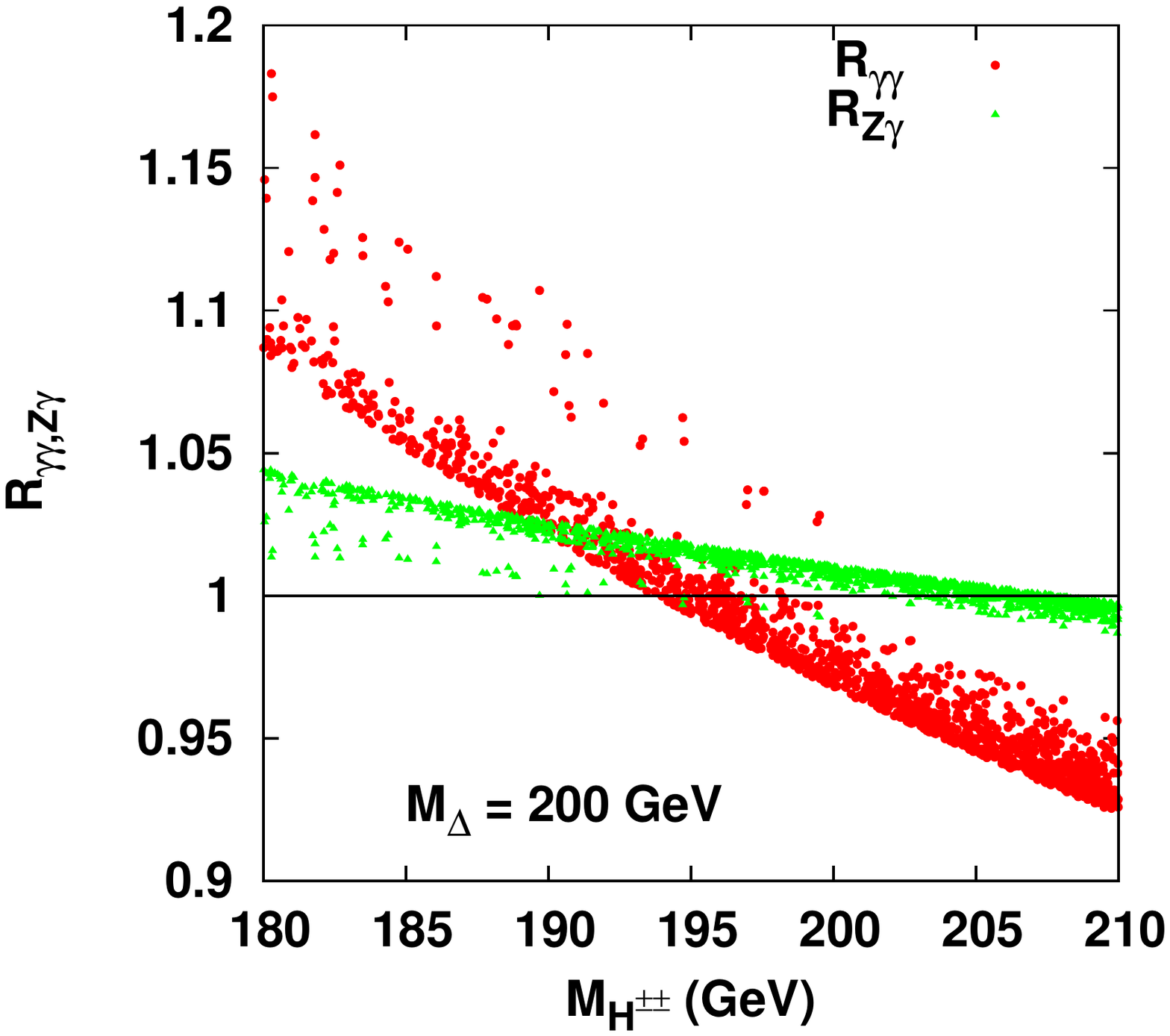}
\hspace{0.5cm}
\includegraphics[width=7cm]{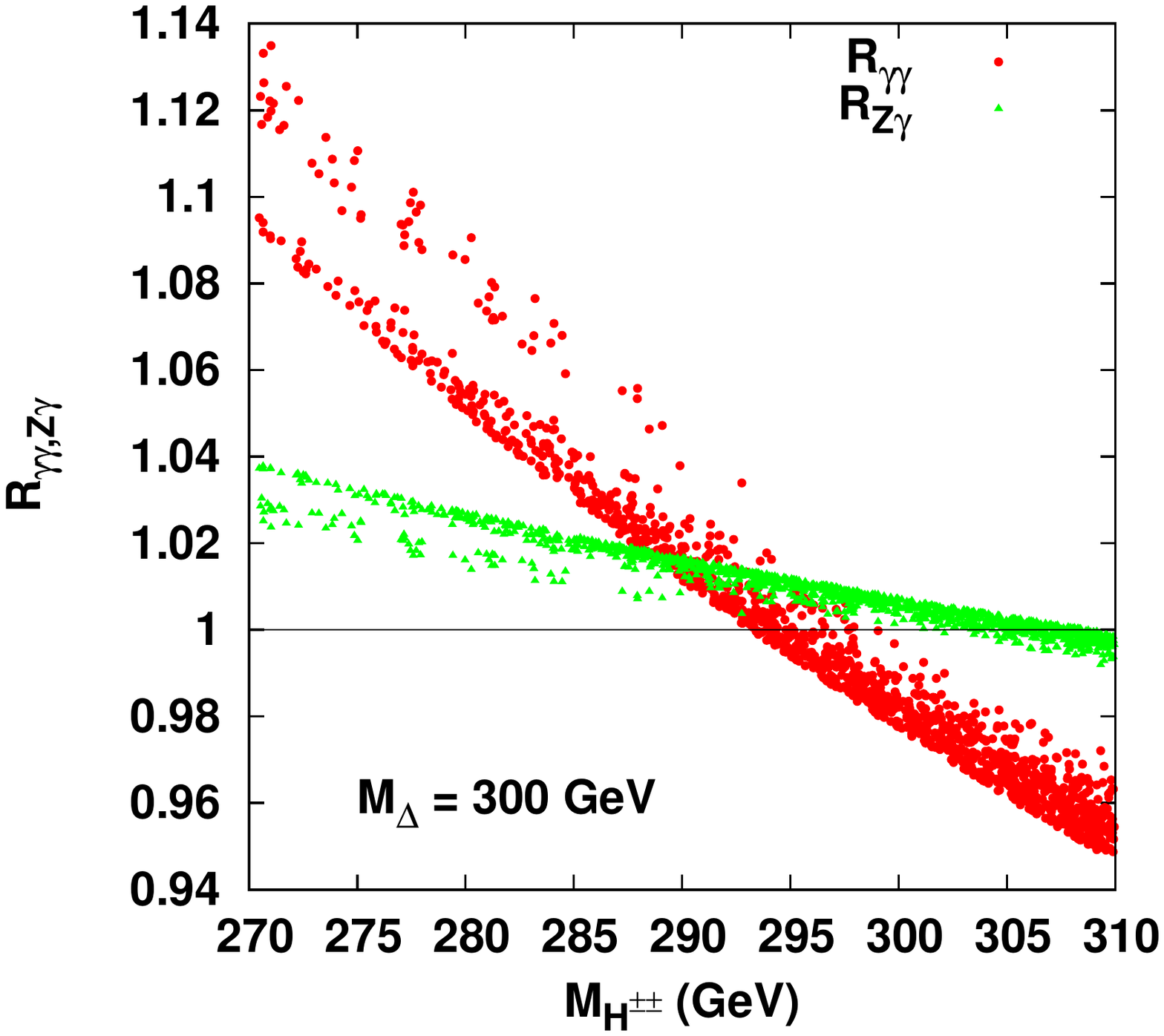}
\vspace{-2.5cm}
\caption{The zoomed-in version of Figure~\ref{fig:Rgg1} to show the 
region of anti-correlation between $R_{\gamma\gamma}$ and $R_{Z\gamma}$.} 
\label{fig:Rggzoom}
\end{figure} 

We note that there exists a small parameter space in which there is an 
{\it anti-correlation} between the $h\to \gamma\gamma$ and $h\to Z\gamma$ 
rates, i.e., $R_{Z\gamma}$ could still be larger than 1 while $R_{\gamma\gamma}$ is below 1, as can be seen for example in Figure~\ref{fig:Rggzoom}. 
However, the enhancement (suppression) of the $Z\gamma~(\gamma\gamma)$ rate in this region is very small of the order of 1\% compared to the SM prediction, and may not be distinguishable at the 
LHC.

It is also worth noting from Fig.~\ref{fig:Rgg1} that the $\gamma\gamma$ 
enhancement can be huge for a light doubly-charged Higgs, and the current 
upper limit from ATLAS, $R_{\gamma\gamma}=1.65^{+0.34}_{-0.30}$~\cite{ATLASnew}, already 
requires $M_{H^{\pm\pm}}>150$ GeV or so. This provides a unique way to probe 
the small Yukawa region ($v_\Delta>10^{-4}$ GeV) of the model parameter space 
which is currently inaccessible to the direct searches. 

\begin{figure}
\centering
\includegraphics[width=7cm]{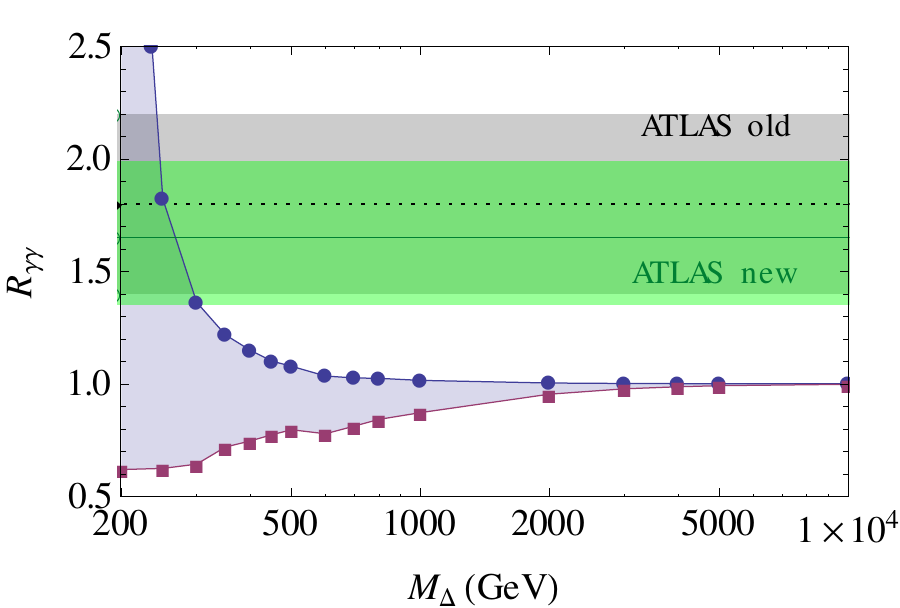} 
\hspace{0.5cm}
\includegraphics[width=7cm]{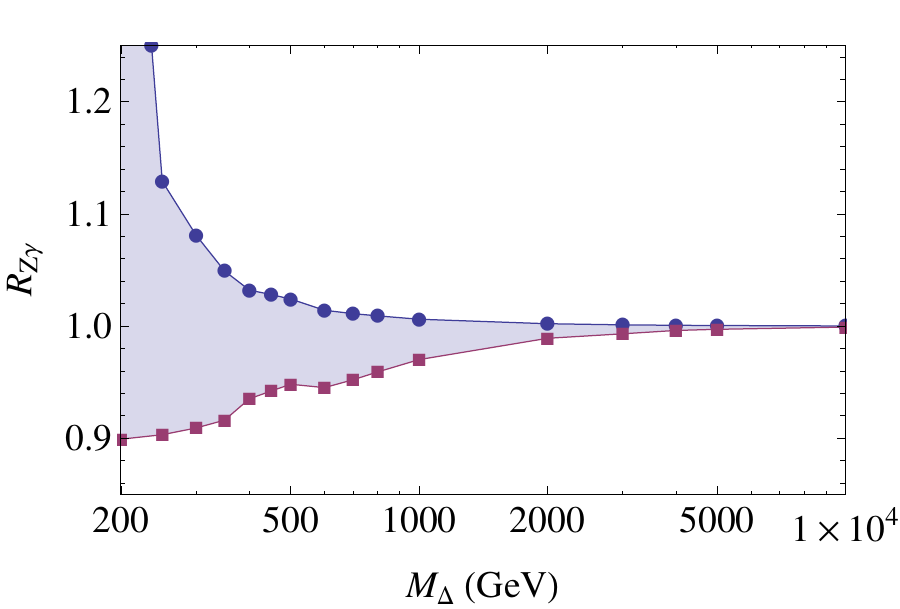}
\caption{The shaded regions show the predicted range of $h\to \gamma\gamma$ and 
$h\to Z\gamma$ decay rates with respect to the SM expectations as a 
function of the type-II seesaw scale for the allowed 
range of the model parameters. The data points denoted 
by circles (squares) 
correspond to the maximum (minimum) value allowed for a given seesaw scale. The dotted horizontal line and the black shaded region show the central value and $1\sigma$ range of the upper limit on $R_{\gamma\gamma}$ from ATLAS~\cite{atlas-gg}. These limits were recently modified~\cite{ATLASnew} to the values shown as green (solid) line and green (dark) shaded region respectively.}\label{fig:upper}
\end{figure}

With the increase in seesaw scale, the enhancement in the 
$\gamma\gamma$ rate 
decreases substantially. In order to quantify this effect, we study the allowed range of $h\to \gamma\gamma,~Z\gamma$ decay rates as a function of the seesaw scale, and as a consequence, obtain an {\it upper} limit on the seesaw scale for any given value of the enhancement. This is explicitly shown in Fig.~\ref{fig:upper}. We find that the new contributions from the charged Higgs bosons in the type-II seesaw model become negligible beyond a seesaw scale of $M_\Delta\sim 10$ TeV. Hence, for a given enhancement in the $\gamma\gamma$ channel, we will have an upper limit on the seesaw scale. For instance, for the current ATLAS central value of $R_{\gamma\gamma}=1.65$ (solid horizontal line in Fig.~\ref{fig:upper}), the upper limit on the seesaw scale becomes 
$M_\Delta<270$ GeV. If more than 10\% enhancement is confirmed in 
future with more statistics, we must have the seesaw scale below 450 GeV 
provided there is no additional contribution to this enhancement due to any 
other new physics effects apart from the type-II seesaw. 

An upper limit on the seesaw scale will in turn put upper bounds on the masses of the singly- and doubly-charged Higgs bosons in the type-II seesaw model which can be used to explore the {\it full} allowed parameter space of the model at the LHC. 
In Tables~\ref{Rgg} and \ref{Rzg}, we have shown the masses of the 
singly- and doubly-charged scalars corresponding to the maximum enhancement (suppression) at  
a particular seesaw scale. This result could be used to test the minimal 
type-II seesaw model as a single viable extension of the SM up to the 
Planck scale, once more precise measurements of the $h\to \gamma\gamma$ rate is 
performed at the LHC. This can also be done in combination with the $h\to Z\gamma$ channel which has currently a poor sensitivity at the LHC~\cite{cms-zg, ATLAS-zg}, but is expected to be improved significantly with more data. A statistically significant enhancement in both the $\gamma\gamma$ and $Z\gamma$ channels would be favorable for a type-II seesaw model.  A detailed study of the correlation between the $\gamma\gamma (Z)$ decay rates and the collider signals of the type-II seesaw model in the allowed parameter space 
will be presented in a future communication.  

\begin{table}
\begin{center}
\begin{tabular}{| c | c | c | c |} 
\hline
$M_\Delta$ [GeV] &  $(R_{\gamma\gamma})_{\rm max(min)}$ & $M_{H^\pm}$ [GeV] & $M_{H^{\pm\pm}}$ [GeV] \\ \hline
200 & 4.41 (0.62)  & 85.02 (480.97) & 100.78 (485.15)  \\ \hline
250 & 1.82 (0.62)  & 165.38 (569.95) & 166.25 (593.44) \\ \hline
300 & 1.36 (0.64) & 236.21 (610.15) & 234.60 (635.74)  \\ \hline
350 & 1.22 (0.72) & 297.15 (576.91) & 295.86 (555.70)  \\ \hline
400 & 1.15 (0.75) & 354.68 (553.48) & 353.60 (592.33)  \\ \hline
450 & 1.10 (0.77) & 415.07 (586.91) & 411.75 (625.70)  \\ \hline
500 & 1.08 (0.80) & 468.82 (626.80) & 465.87 (665.15)  \\ \hline
600 &  1.03 (0.78) & 580.43 (799.78) & 580.80 (820.96) \\ \hline
700 &  1.03 (0.81) & 681.00 (866.34) & 683.21 (901.48) \\ \hline
800 & 1.02 (0.83) & 782.18 (963.23) & 781.28 (992.81)  \\ \hline
1000 & 1.02 (0.87) & 985.80 (1129.40) & 985.08 (1168.71) \\ \hline
2000 & 1.00 (0.95) & 1993.26 (2070.82) & 1992.46 (2105.19) \\ \hline
3000 & 1.00 (0.98) & 2995.54 (3047.69) & 2994.98 (3071.13) \\ \hline
4000 & 1.00 (0.99) & 3996.68 (4035.92) & 3996.23 (4053.62) \\ \hline
5000 & 1.00 (0.99) & 4997.38 (5028.82) & 4996.99 (5042.99) \\ \hline
$10^4$ - $10^{10}$ & 1.00 & ${\cal O}(M_\Delta)$ & ${\cal O}(M_\Delta)$  \\ \hline  
\end{tabular}
\end{center} 
\caption{Masses of the charged scalar particles in the type-II seesaw model corresponding to the maximum (minimum) value of $R_{\gamma\gamma}$ for a given 
seesaw scale, as shown in Fig.~6.}\label{Rgg}
\end{table}

\begin{table}
\begin{center}
\begin{tabular}{| c | c | c | c |} 
\hline
$M_\Delta$ [GeV] &  $(R_{Z\gamma})_{\rm max(min)}$ & $M_{H^\pm}$ [GeV] & $M_{H^{\pm\pm}}$ [GeV] \\ \hline
200 &  1.59 (0.90) & 124.53 (502.52) & 101.47 (538.36) \\ \hline
250 & 1.13 (0.90) & 177.62 (587.18) & 177.88 (613.73) \\ \hline
300 &  1.08 (0.91) & 236.20 (599.43) & 234.59 (631.45) \\ \hline
350 & 1.05 (0.92) & 298.90 (625.95) & 296.22 (656.72) \\ \hline
400 &  1.03 (0.93) & 365.43 (551.23) & 357.14 (589.91) \\ \hline
450 &  1.03 (0.94) & 452.68 (586.91) & 414.06 (625.70) \\ \hline
500 &  1.02 (0.95) & 501.12 (623.13) & 464.28 (662.72) \\ \hline
600 &  1.01 (0.95) & 614.53(799.78) & 576.04 (820.96) \\ \hline
700 &  1.01 (0.95) & 713.08 (861.61) & 677.86 (899.68) \\ \hline
800 &  1.01 (0.96) & 810.44 (940.12) & 780.25 (980.01) \\ \hline
1000 & 1.01 (0.97) & 1020.84 (1125.07) & 985.97 (1150.29) \\ \hline
2000 & 1.00 (0.99) & 2027.84 (2063.83) & 1993.79 (2100.74) \\ \hline
3000 & 1.00 (0.99)& 3037.13 (3039.13) & 2997.66 (3078.23) \\ \hline
4000 & 1.00 (1.00) & 4027.95 (4031.07) & 3998.24 (4061.69) \\ \hline
5000 & 1.00 (1.00) & 5022.42 (5027.66) &  4998.59 (5053.46) \\ \hline
 \hline
$10^4 - 10^{10}$ & 1.00 & ${\cal O}(M_\Delta)$   & ${\cal O}(M_\Delta)$  \\ 
 \hline  
\end{tabular}
\end{center} 
\caption{Masses of the charged scalar particles in the type-II seesaw model corresponding to the maximum (minimum) value of $R_{Z\gamma}$ for a given seesaw scale, as shown in Fig.~6.}\label{Rzg}
\end{table}

\section{Conclusions}\label{6}
In conclusion, we have studied the effects of electroweak vacuum stability and unitarity conditions on the full parameter space of the minimal type-II seesaw model in the light of the recent discovery of a SM Higgs-like particle at the LHC in the mass range of 124 - 126 GeV. We find that there exists a large parameter space in the model, irrespective of the seesaw scale (as long as it is below the SM vacuum instability scale), which yields a SM-like Higgs mass around 125 GeV while satisfying all the stability and unitarity conditions as well as neutrino oscillation data, collider bounds and other low-energy data. We have performed  a numerical scan over the entire parameter space of the model and have shown the allowed region for two benchmark values representing low- and high-scale 
seesaw. We have also studied the predictions for the partial decay widths of the $h\to \gamma\gamma$ and $h\to Z\gamma$ with respect to their SM expectations and find that these two rates are correlated in the type-II seesaw model. Moreover, for a sufficiently low seesaw scale, the deviations from the SM prediction could be significant. For a given value of such deviation, we obtain an upper bound on the seesaw scale. This in turn imposes an upper bound on the masses of the singly- and doubly-charged Higgs bosons in the model. For more than 10\% deviation of the $\gamma\gamma$ signal strength from its SM value, the 
corresponding upper bound on the type-II seesaw scale is about 450 GeV which is completely within the reach of the LHC. This result should be encouraging for the experimental searches for these charged Higgs particles. With improved 
sensitivity in the $h\to \gamma\gamma$ and $Z\gamma$ signal strengths in the 
future, we hope to be able to probe the entire allowed parameter space of the 
minimal type-II seesaw model, thus enabling us to firmly establish/eliminate it as a single viable extension of the SM. 
\section*{Acknowledgments} 
We thank Goran Senjanovi\'{c} for his valuable input. PSBD acknowledges the local hospitality provided at the Indian Association for the Cultivation of Science (IACS), Kolkata, 
where this work was initiated, and would like to thank Csaba Bal\'{a}zs, Abdelhak Djouadi and Anupam Mazumdar for useful discussions. 
The work of PSBD is supported by the 
Lancaster-Manchester-Sheffield Consortium for Fundamental Physics under STFC 
grant ST/J000418/1. DKG would like to acknowledge the hospitality provided by 
the University of Helsinki and the Helsinki Institute of Physics where part
of this work was done. The work of NO is supported in part by the DOE Grant No. DE-FG02-10ER41714. IS would like to thank Anirban Dutta for helpful 
discussions.

\appendix 
\section{Matching condition for the $\overline{\rm MS}$ and pole masses}
 \label{A} 
The running Higgs mass in the $\overline{\rm MS}$ scheme is related to its pole mass by the matching condition given by Eq.~(\ref{match-mh}), where~\cite{sirlin} 
\beq
 \Delta_h(\mu) = \frac{G_F}{\sqrt{2}} \frac{M_Z^2}{8 \pi^2}
 \left[
   \xi f_1\left(\xi,\mu \right)
 + f_0\left(\xi,\mu \right)
 + \frac{1}{\xi} f_{-1}\left(\xi,\mu \right)
 \right],
\label{delta-h}
\eeq
with $\xi\equiv M_h^2/M_Z^2$. The loop-functions $f(\xi)$ are given by
\beq
f_1(\xi,\mu) &=& 6\ln\left(\frac{\mu^2}{M_h^2}\right)
  + \frac{3}{2}  \ln \xi - \frac{1}{2} {\cal Z}\left(\frac{1}{\xi}\right)
 -{\cal Z}\left(\frac{c_W^2}{\xi}\right) -\ln c_W^2
  +\frac{9}{2} \left( \frac{25}{9} - \frac{\pi}{\sqrt{3}}
  \right), \nonumber\\
f_0(\xi)  &=& - 6\ln\left(\frac{\mu^2}{M_Z^2}\right)
  \left[ 1 +2 c_W^2 -2 \frac{M_t^2}{M_Z^2} \right]
 +\frac{3 c_w^2 \xi}{\xi-c_W^2} \ln \frac{\xi}{c_W^2}
  +2 {\cal Z}\left( \frac{1}{\xi} \right) 
+ 4 c_W^2 {\cal Z}\left( \frac{c_W^2}{\xi} \right)\nonumber\\
 &&  +\left(\frac{3 c_W^2}{s_W^2} +12 c_W^2\right) \ln c_W^2
  -\frac{15}{2} \left( 1 +2 c_W^2 \right) 
- 3\frac{M_t^2}{M_Z^2} \left[
     2 {\cal Z}\left( \frac{M_t^2}{M_Z^2 \xi} \right)
    +4 \ln \frac{M_t^2}{M_Z^2} -5 \right], \nonumber\\
f_{-1}(\xi) &=& 6 \ln\left( \frac{\mu^2}{M_Z^2}\right)
  \left[ 1 +2 c_W^4 -4\frac{M_t^4}{M_Z^4} \right]
 -6 {\cal Z}\left( \frac{1}{\xi} \right)-12 c_W^4 {\cal Z}\left( \frac{c_W^2}{\xi} \right)
 -12 c_W^4 \ln c_W^2 \nonumber\\
 && + 8\left( 1 +2 c_W^4 \right) +24 \frac{M_t^4}{M_Z^4}
  \left[\ln \frac{M_t^2}{M_Z^2} -2 + {\cal Z}\left( \frac{M_t^2}{M_Z^2 \xi} \right)
\right],
\eeq
with $s_W^2 \equiv \sin^2 \theta_W$, $ c_W^2 \equiv \cos^2 \theta_W$
 ($\theta_W$ denotes the weak mixing angle) and
\beq
 {\cal Z}(z) = \left\{
  \begin{array}{cc}
    2 {\cal A} \tan^{-1}(1/{\cal A})\, & (z > 1/4 ) \\
    {\cal A} \ln \left[ (1+{\cal A})/(1-{\cal A}) \right]\, & (z < 1/4 ) ,
 \end{array}\right.
\eeq
with ${\cal A} = \sqrt{ \left| 1 - 4 z \right| }$. 

For the matching condition (\ref{match-mt}) between the top quark $\overline{\rm MS}$ and pole masses, the radiative correction has both QCD~\cite{gray, Fleischer, chety, melnikov} and electroweak~\cite{Hempfling, kalmy} parts, each of which is separately finite and gauge-independent. Here we give the explicit expression up to ${\cal O}(\alpha_3^2)$ for the QCD part and ${\cal O}(\alpha)$ for the electroweak part:  
\begin{eqnarray}
\Delta_t(\mu) &=& \left[\ln\left(\frac{M_t^2}{\mu^2}\right)-\frac{4}{3}\right]\left(\frac{\alpha_3(\mu)}{\pi}\right)
+ (1.0414 N_L-14.3323)\left(\frac{\alpha_3(\mu)}{\pi}\right)^2\nonumber\\
&&
+\frac{1}{3}\left[\ln\left(\frac{M_t^2}{\mu^2}\right)-\frac{4}{3}\right]\left(\frac{\alpha(\mu)}{\pi}\right)\nonumber\\
&& +\frac{G_F}{\sqrt 2} \frac{M_t^2}{8\pi^2}
\left[-\frac{9}{2}\ln\left(\frac{M_t^2}{\mu^2}\right)+\frac{11}{2}-r+2r(2r-3)\ln(4r)-8r^2\left(\frac{1}{r}-1\right)^{3/2}\cos^{-1}(\sqrt r)\right]\nonumber\\
&& +a_t+b_t\ln\left(\frac{M_h}{300~{\rm GeV}}\right)+c_t\ln\left(\frac{M_t}{175~{\rm GeV}}\right)
\label{delta-mt}
\end{eqnarray}
where $N_L$ is the number of massless quark flavors, $r\equiv M_h^2/4M_t^2$. For $\mu=M_t$, the numerical coefficients $(a_t,b_t,c_t)=
(-6.90,1.73,-5.82)\times 10^{-3}$~\cite{Hempfling,Schrempp}. We have neglected 
the ${\cal O}(\alpha\alpha_3^2)$ and ${\cal O}(\alpha_3^3)$ terms in 
Eq.~(\ref{delta-mt}) whose contributions are less than 0.5\%~\cite{kalmy}.

Following the latest best-fit mass measurement results, $M_h=125.2\pm 0.3 ({\rm stat})\pm 0.6 ({\rm syst})$ GeV (ATLAS)~\cite{atlas-dec12} and $125.8\pm 0.4({\rm stat}) \pm 0.4 ({\rm syst})$ GeV (CMS)~\cite{cms-dec12}, we have used $M_h=125$ GeV in Eq.~(\ref{match-mh}) and for the rest of our analysis, unless 
otherwise specified. 

For the top quark pole mass, we have used $M_t=173.2$ GeV in Eq.~(\ref{match-mt}) following the latest measurements from Tevatron and LHC experiments which are consistent with each other: $M_t=173.2\pm 0.9$ GeV (Tevatron)~\cite{tevtop} and $173.3\pm 0.5\pm 1.3$ GeV (ATLAS+CMS)~\cite{LHC-top}. For the other SM 
parameters appearing in Eqs.~(\ref{delta-h}) and (\ref{delta-mt}), we have used the PDG central values: $G_F=1.166\times 10^{-5}~{\rm GeV}^{-2}$ for the Fermi coupling constant, $\alpha(M_t)=1/127.9$ for the fine-structure 
constant, $M_W=80.4$ GeV and $M_Z=91.2$ GeV for the $W$ and $Z$ pole masses~\cite{pdg}. 

\section*{Note added}
We found that the amplitude of the scalar triplet contribution in the type-II seesaw model to the partial decay width of $h\to Z\gamma$  given in the previous version of our paper (following Ref.~\cite{carena, frank}) had a wrong relative sign as compared to the 
SM contribution, as also pointed out recently in Ref.~\cite{Chen:2013vi}. 
This leads to significantly 
different predictions for the $h\to Z\gamma$ rate and its correlation with the $h\to \gamma\gamma$ rate than those presented earlier. 
In this version, we correct the expression for the partial decay width of 
$h\to Z\gamma$ given in Eq. (\ref{eq:hzg}), and 
show that it is {\it correlated} with the $h\to\gamma\gamma$ partial width over most of the allowed parameter space (see Fig.~\ref{fig:Rgg1}). Similar conclusions were derived independently in Ref.~\cite{Chen:2013dh}.

\end{document}